\documentclass[a4paper,clock]{article}
\usepackage[dvips]{epsfig}
\usepackage{amssymb}
\usepackage{bm}
\usepackage{dcolumn}
\usepackage{amsmath}

\catcode`\@=11
\def\marginnote#1{}
\newcount\hour
\newcount\minute
\newtoks\amorpm
\hour=\time\divide\hour by60
\minute=\time{\multiply\hour by60 \global\advance\minute
by-\hour}\edef\standardtime{{\ifnum\hour<12
\global\amorpm={am}%
        \else\global\amorpm={pm}\advance\hour by-12 \fi
        \ifnum\hour=0 \hour=12 \fi
        \number\hour:\ifnum\minute<10
0\fi\number\minute\the\amorpm}}
\edef\militarytime{\number\hour:\ifnum\minute<10
0\fi\number\minute}

\def\draftlabel#1{{\@bsphack\if@filesw {\let\thepage\relax
   \xdef\@gtempa{\write\@auxout{\string
      \newlabel{#1}{{\@currentlabel}{\thepage}}}}}\@gtempa
   \if@nobreak \ifvmode\nobreak\fi\fi\fi\@esphack}
        \gdef\@eqnlabel{#1}}
\def\@eqnlabel{}
\def\@vacuum{}
\def\draftmarginnote#1{\marginpar{\raggedright\scriptsize\tt#1}}
\def\draft{\oddsidemargin -.5truein
        \def\@oddfoot{\sl preliminary draft \hfil
        \rm\thepage\hfil\sl\today\quad\militarytime}
        \let\@evenfoot\@oddfoot \overfullrule 3pt
        \let\label=\draftlabel
        \let\marginnote=\draftmarginnote

\def\@eqnnum{(\theequation)\rlap{\kern\marginparsep\tt\@eqnlabel}%
\global\let\@eqnlabel\@vacuum}  }

\def\numberbysection{\@addtoreset{equation}{section}
        \def\theequation{\thesection.\arabic{equation}}}

\def\underline#1{\relax\ifmmode\@@underline#1\else
 $\@@underline{\hbox{#1}}$\relax\fi}

\catcode`@=12
\relax

\numberbysection

\advance \topmargin by -\headheight
\advance \topmargin by -\headsep

\evensidemargin \oddsidemargin
\def\rf#1{(\ref{#1})}
\def\lab#1{\label{#1}}

\def\br{\begin{eqnarray}}
\def\er{\end{eqnarray}}
\def\be{\begin{equation}}
\def\ee{\end{equation}}

\def\({\left(}
\def\){\right)}

\newcommand{\bi}[1]{\bibitem{#1}}

\relax

\def\a{\alpha}

\def\b{\beta}

\def\d{\delta}

\def\g{\gamma}

\def\pa{\partial}

\def\tp0{\Theta_{+}^{(0)}}
\def\tm0{\Theta_{-}^{(0)}}

\def\vp{\varphi}

\def\f#1#2#3 {f^{#1#2}_{#3}}

\def\win1{{\sf w_{1+\infty}}}

\def\Win1{{\sf W_{1+\infty}}}

\def\rlx{\relax\leavevmode}
\def\inbar{\vrule height1.5ex width.4pt depth0pt}
\def\IZ{\rlx\hbox{\sf Z\kern-.4em Z}}
\def\IR{\rlx\hbox{\rm I\kern-.18em R}}
\def\IC{\rlx\hbox{\,$\inbar\kern-.3em{\rm C}$}}
\def\IN{\rlx\hbox{\rm I\kern-.18em N}}
\def\IO{\rlx\hbox{\,$\inbar\kern-.3em{\rm O}$}}
\def\IP{\rlx\hbox{\rm I\kern-.18em P}}
\def\IQ{\rlx\hbox{\,$\inbar\kern-.3em{\rm Q}$}}
\def\IF{\rlx\hbox{\rm I\kern-.18em F}}
\def\IG{\rlx\hbox{\,$\inbar\kern-.3em{\rm G}$}}
\def\IH{\rlx\hbox{\rm I\kern-.18em H}}
\def\II{\rlx\hbox{\rm I\kern-.18em I}}
\def\IK{\rlx\hbox{\rm I\kern-.18em K}}
\def\IL{\rlx\hbox{\rm I\kern-.18em L}}
\def\one{\hbox{{1}\kern-.25em\hbox{l}}}
\def\0#1{\relax\ifmmode\mathaccent"7017{#1}%
B        \else\accent23#1\relax\fi}

\def\NPB#1#2#3{{\sl Nucl. Phys.} {\bf B#1} (#2) #3}

\def\PRD#1#2#3{{\sl Phys. Rev.} {\bf D#1} (#2) #3}

\def\PLB#1#2#3{{\sl Phys. Lett.} {\bf #1B} (#2) #3}

\def\JPAMT#1#2#3{{\sl J. Physics A: Math. Theor.} {\bf A#1} (#2) #3}

\def\JHEP#1#2#3{{\sl JHEP} {\bf #1} (#2) #3}

\def\JPAMG#1#2#3{{\sl J. Physics A: Math. Gen.} {\bf A#1} (#2) #3}

\def\UN#1#2#3{{\sl Universe} {\bf #1} (#2) #3}
\def\EPJC#1#2#3{{\sl Eur. Phys. J. C} {\bf #1} (#2) #3}
 
\def\MoC#1#2#3{{\sl MATHEMATICS OF COMPUTATION} {\bf #1} (#2) #3}
\def\ChPB#1#2#3{{\sl Chinese Phys. B} {\bf #1} (#2) #3}
 
\begin{document}
\begin{titlepage}

\vskip .6in

\begin{center}
{\large {\bf Hirota–tau and Heun-function framework for Dirac vacuum polarization and quantum stabilization of kinks}}
\end{center}

\normalsize
\vskip .4in

\begin{center}

Harold Blas 

\par \vskip .2in \noindent

$^{a}$ Instituto de F\'{\i}sica\\
Universidade Federal de Mato Grosso\\
Av. Fernando Correa, $N^{0}$ \, 2367\\
Bairro Boa Esperan\c ca, Cep 78060-900, Cuiab\'a - MT - Brazil. \\ 
\normalsize
\end{center}
\par \vskip .3in \noindent

We investigate a modified affine Toda model coupled to matter (ATM) which includes a scalar self-interacting potential and demonstrate that its first-order integro-differential structure, preserving a deformed Noether-topological current correspondence, provides a consistent framework for fermion-soliton interactions. In this formulation, the fermion-soliton energy is proportional to the soliton’s topological charge. We evaluate the renormalized energy functional, incorporating one-loop quantum corrections, and perform a variational minimization to determine the configuration that extremizes the functional. The fermionic back-reaction and the self-interacting scalar critically shape the fermion-kink energy, the in-gap bound-state spectrum, and the fermionic vacuum-polarization energy, yielding well-defined stability minima of the total energy as functions of the fermion and scalar masses and coupling parameters, in the semiclassical approximation. A key result is that the Heun-equation formalism is necessary to construct nonzero-energy bound and scattering states: unlike the tau-function method, which captures only the zero mode, the Heun approach encodes the full scattering data through local solution matching conditions. These results refine the spectral analysis of deformed integrable models. The stability of soliton-fermion configurations has direct implications for topologically protected states in quantum information and condensed-matter systems.

\end{titlepage}

\section{Introduction}

Integrable models are central to theoretical physics, providing a framework to analyze complex classical and quantum dynamics \cite{raja, babelon, frishman, abdalla}. In particular, $sl(n)$ affine Toda models coupled to matter (ATM) offer a versatile setting to study the interplay between bosonic and fermionic fields. By extending the traditional Toda model to include matter couplings, ATM models capture a wide range of phenomena, including nonlinearity, topological defects, chiral confinement, bound states, and the correspondence between Noether and topological charges. Their capacity to describe soliton–fermion configurations and associated nonlinear and topological effects makes them a powerful tool for probing integrable and quasi-integrable dynamics for Hermitian and non-Hermitian systems \cite{matter, npb1, npb2, prd, jhep22, jhep24, prd2}.

Fermion back-reaction on kinks is a topic of ongoing interest, with important implications for non-perturbative phenomena in quantum field theory. Kink–fermion systems typically feature a fermion zero mode and charge fractionalization \cite{jackiw1}, while higher-energy valence states may arise as excitations of the bound spectrum. Recent studies have mainly relied on numerical constructions of kink configurations including fermion back-reaction or on analyses assuming a predetermined kink background \cite{shnir1, shnir2, gani, mohammadi, mohammadi1}. The total energy of a fermion–kink system consists of the classical fermion–soliton interaction energy, the bound-state fermion energies, and the fermionic vacuum-polarization energy (VPE). The VPE, stemming from the interaction with the Dirac sea, is crucial for the consistency of semiclassical expansions and for understanding how fermionic back-reaction influences kink stability and dynamics \cite{weigel1, weigel2, prd2}.

Due to the challenge of obtaining exact analytical results in general models, we study a deformation of an integrable system to probe fermion back-reaction. Analytical solutions are obtained via a hybrid Hirota–tau and Heun-equation approach, allowing a systematic analysis of kink profiles, fermionic bound states, and scattering states as functions of model parameters. Our results demonstrate that the back-reaction of both localized and scattering fermions plays a decisive role in shaping the system’s spectra.

We study a fermion–soliton system with fixed topological charge $\frac{1}{2}$, realized as a solution to a set of first-order integro-differential equations. A related model, in which quantum effects can stabilize solitons, was analyzed in \cite{farhi, farhi2}. The present model may be regarded as a specific reduction of that framework when their scalar fields $\phi_{1,2}$ are constrained to the chiral circle, yielding a sine-Gordon–type self-interaction potential. In this regime, our fermion–soliton configurations correspond to classical solutions of the model in \cite{farhi, farhi2} for appropriate choices of their parameter space. In fact, when the scalar field is constrained to lie on the chiral circle, the model becomes highly specialized. This constraint significantly simplifies the analysis by eliminating the need for the otherwise complex procedures of regularization and renormalization.
 
Our analysis of the modified ATM model uncovers several structural features that enhance the understanding of kink–fermion systems. Departing from the standard Bogomolnyi construction, we employ the framework of \cite{prd2}, wherein first-order equations follow from the equivalence between Noether and topological charges. In this formulation, the fermion–soliton energy becomes proportional to the soliton’s topological charge.

In semiclassical analyses of kink solitons coupled to excited fermionic bound states, both the bound-state energy and the Dirac-sea contribution must be treated on equal footing \cite{weigel1, weigel2, prd2}. In this work, the Dirac-sea contribution is evaluated as the fermionic vacuum-polarization energy (VPE) and incorporated together with the fermion–soliton interaction energy and the bound-state energy into the total energy functional. Our approach differs from the exact tau-function method of \cite{prd2}, which yields scattering phase shifts analytically; here, phase shifts are obtained from matching conditions of the scattering states within the local Heun-function formalism. Consequently, the computation of the VPE requires a combination of analytical and numerical techniques.

The Heun-equation is a subject of active research, partly because the band structure of its solutions plays a key role in the theory of integrable nonlinear wave equations (see \cite{maier} and references therein). Heun’s equation has also emerged as a key tool in contemporary theoretical and mathematical physics, providing a unifying framework for analyzing systems whose complexity exceeds the scope of simpler special functions. As the most general second-order linear ODE with four regular singular points, extending the hypergeometric equation with its three, it naturally arises in problems involving intricate potentials, boundary conditions, or geometries \cite{ronveaux, slavyanov, hortacsu, olver}.
  
The role of the Heun-equation framework in the present work warrants particular emphasis, as it complements the Hirota-tau formalism by enabling the explicit determination of both nonzero bound states and scattering states. Our model coincides with that of Ref. \cite{loginov} within the sector involving the diagonal spinor components $(u, v)^T$ of their  $2\times 2$
 spin–isospin matrix coupled to external sine-Gordon soliton. However, whereas Ref. \cite{loginov} treats the sine-Gordon soliton as an external, prescribed background, the present work incorporates a fully dynamical soliton and systematically accounts for the backreaction induced by the fermionic sector.
 
Following the approach of \cite{farhi,farhi2}, we determine the minimum–energy configuration at fixed fermion number using a parametrized variational ansatz for the kink profile. The variational parameter is the inverse soliton width, $2 K$. The exact self-consistent soliton solution is expected to lie close to this variational minimum and to possess a lower energy. Consequently, the soliton energy satisfies $E_{tot} < M$\, ($M$ being the  fermion mass), implying absolute stability.

The paper is organized as follows. Section \ref{sec:model} introduces the model and its associated first-order integro-differential equations. Section \ref{sec:bs} derives the soliton–fermion configurations and the spinor zero modes. In Section \ref{sec:energy}, we compute the energy of kink–fermion configurations including the fermionic bound states. Section \ref{sec:Heun} analyzes the scattering and nonzero-energy and zero-mode bound states using the Heun-equation formalism. The fermionic vacuum-polarization energy (VPE) is evaluated in Section \ref{sec:vpe}. Section \ref{sec:totalenergy} presents the total energy and the corresponding stability points in parameter space. Finally, Section \ref{sec:discuss} contains the discussion and concluding remarks. Appendix A presents the spectral symmetry and in Appendix B it is shown the reciprocal left/right scattering property of the model.

\section{The model}
\label{sec:model}
We consider the field theory in $1+1$ dimensions defined by the Lagrangian\footnote{Our notation: $x_{\pm} = t\pm x $, and so, $\pa_{\pm}=\frac{1}{2} (\pa_t \pm \pa_x)$, and $\pa^2=\pa^2_t -\pa^2_x = 4\pa_{-}\pa_{+}$. We use $\g_0 = \(\begin{array}{cc} 0  & i\\
-i & 0\end{array}\)$, $\g_1 = \(\begin{array}{cc} 0  & -i\\
-i & 0\end{array}\)$, $\g_5 = \g_0 \g_1 = \(\begin{array}{cc} 1  & 0\\
0 & -1\end{array}\)$,  and $\psi = \(\begin{array}{c} \psi_{R}\\
\psi_{L}\end{array}\),\,\, \bar{\psi} = \psi^{\dagger} \g_0,\,  \psi_{R} \equiv (\frac{1+\g_5}{2})\psi,\, \psi_{L} \equiv (\frac{1-\g_5}{2})\psi $.} 
\br
\lab{atm0}
{\cal L} =\frac{1}{2}\partial_{\mu }\varphi \partial ^{\mu }\varphi +i\overline{\psi }\gamma ^{\mu}\partial _{\mu }\psi - M \overline{\psi }e^{2 i \hat{\b} \varphi
\gamma _{5}}\psi - A_1 (1-\cos{(2 \hat{\b} \vp)}), 
\er
where $\vp$ is a real scalar field, $\psi$ is a Dirac spinor, $M$ and $A_{1}$ are real parameters and $\hat{\b}$ is the coupling constant. The term of the scalar self-coupling potential is new and constitutes a relevant modification of the $sl(2)$ affine Toda system coupled to matter field (ATM). A family of ATM integrable models have been discussed in \cite{matter, npb2}. The modified model (\ref{atm0}) is not integrable; however, some techniques  of  integrable systems can be used in this context, such as the construction of kinks and spinor bound states in the tau function approach. In (\ref{atm0}) we consider $N_f$ Dirac fermions chirally coupled to $\vp$. For notational simplicity, the flavor index is omitted throughout, while overall factors of $N_f$ are retained where relevant. Technically, $N_f$ flavors are introduced as a control parameter for a large-$N_f$ expansion, in which fermion-loop effects dominate and stabilize the soliton, while bosonic and higher-order corrections are parametrically suppressed. 

The ATM model - corresponding to Eq. (\ref{atm0}) with 
$A_1=0$ - has recently been analyzed in the context of fermionic backreaction on kink configurations and a topological charge-pumping mechanism, incorporating the classical fermion–soliton interaction energy, the bound-state fermion spectrum, and the fermionic vacuum polarization energy \cite{prd2}. However, the topological sector with 
$Q_{top} = \pm \frac{1}{2}$ was not examined in that work, since the scattering states in this regime cannot be constructed within the Hirota–tau formalism adopted there. In the present study, we revisit this sector and show that the realization of stable fermion–soliton configurations necessitates the introduction of a self-interaction potential for the scalar field. To analyze the scattering spectrum and the corresponding vacuum polarization energy, we employ a hybrid analytical–numerical approach based on the Heun-type differential equation.

An analogous model, in which quantum effects provide a stabilization mechanism for solitons, has been analyzed in \cite{farhi, farhi2}. In fact, the model (\ref{atm0}) with scalar self-coupling term ($A_1\neq 0$) becomes a sub-model of the one in \cite{farhi, farhi2} provided that their scalar fields $\phi_{1,2}$ lie on the chiral circle $(\phi_1\,,\,\phi_2) = \frac{1}{2 \hat{\b}} (\cos{(2 \hat{\b} \vp)}\, ,\, \sin{(2 \hat{\b} \vp)})$, where $\vp(x \rightarrow -\infty) \rightarrow 0$ and $\vp(x \rightarrow +\infty) \rightarrow \pi/\hat{\b}$.

In addition, the spinor sector of the ATM model with a prescribed sine-Gordon–type solitonic background has been investigated through numerical methods and the phase-shift formalism to determine the corresponding Casimir energy \cite{mohammadi, mohammadi1}. In contrast, the present work focuses on self-consistent solitonic configurations, wherein the backreaction of the spinor field is fully incorporated into the exact solutions of the model. 

The equations of motion in components become
\br
\label{scalar}
\pa^2_t \vp - \pa^2_x \vp + 2 \hat{\b} M (e^{-2i \hat{\b} \vp} \psi^{\star}_{R}\psi_{L} + e^{2i \hat{\b} \vp} \psi^{\star}_{L}\psi_{R}) +2 \hat{\b} A_1 \sin{(2 \hat{\b} \vp)}  &=&0,
\\
\label{xis11}
(\pa_t + \pa_x) \psi_{L} + M e^{2i \hat{\b} \vp }
 \psi_{R} &=& 0\\
(\pa_t - \pa_x) \psi_{R} -  M e^{-2i \hat{\b} \vp} \, \psi_{L}  &=& 0, \label{xis12}
\er
plus the complex conjugations of the equations (\ref{xis11})-(\ref{xis12}) .

In this work, we examine several distinctive properties of the model at the quasi-classical level, together with its soliton and bound-state solutions. Our analysis employs the Hirota–tau function method in conjunction with the Heun equation formalism to determine the bound-state spectrum and the corresponding scattering solutions. Furthermore, we adopt a hybrid analytical–numerical strategy to compute the energies of the zero modes and valence fermions, as well as the phase shifts of the scattering spinor components. The vacuum polarization energy is evaluated numerically, utilizing analytical results derived from the Wronskian formalism to match two locally analytic solutions of the Heun equation.

\subsection{First order integro-differential equations}

\label{sec:chi}

 Let us consider the two-component spinor parametrized as
\br
\label{spbs1}
\psi   = e^{- i \epsilon t}\(\begin{array}c
\xi_R(x)\\
\xi_L(x)\end{array}\).
\er
So, using (\ref{spbs1}) in the spinor sector (\ref{xis11})-(\ref{xis12}) one can write the coupled system of static equations
\br
\label{sta1}
- i \epsilon \xi_L + \pa_x \xi_L + M e^{2 i \hat{\b} \vp} \xi_R &=&0,\\
\label{sta2}
i \epsilon \xi_R + \pa_x \xi_R + M e^{-2i \hat{\b} \vp} \xi_L &=&0.
\er   
Next, let us consider the integro-differential equation  
\br
\label{topcurr0}
\xi^\star_R \xi_R + \xi^\star_L\xi_L + \frac{1}{\hat{\b}} \pa_x\vp - 2 A_1 \int_{-\infty}^{x} dx \, \sin{(2 \hat{\b} \vp)}  = 0. 
\er
 
Remarkably, one can check that the first order integro-differential equation (\ref{topcurr0}) together with the first order system of equations  (\ref{sta1})-(\ref{sta2}) reproduces the static version of the second order equation (\ref{scalar}). This happens for any value of the parameter $\epsilon$ in (\ref{sta1})-(\ref{sta2}), i.e. for the zero-modes and the fermionic excited states.  So, one expects that the solutions of the first order system of integro-differential eqs.  (\ref{topcurr0})  and (\ref{sta1})-(\ref{sta2}) will solve the second order differential eq.  \rf{scalar} for the scalar field  $\vp$.

In various nonlinear field theories, relevant solutions can be obtained by reducing the Euler–Lagrange equations to first-order systems, such as Bogomolny, Bäcklund, or self-duality equations, thereby enhancing analytical tractability and revealing structural features of the theory \cite{adam1, ferreira1}.

In our case, as we will show  below the first-order integro-differential system (\ref{topcurr0}) and (\ref{sta1})–(\ref{sta2}) provides a more tractable framework for obtaining the soliton and bound-state solutions of the model (\ref{atm0}), as well as a closed form of the energy in terms of the fractional topological charges of the soliton $Q_{k(\bar{k})} = \pm  \frac{1}{2}$, the coupling constant $\hat{\b}$ and the masses $M$ and $m\, (m \equiv  4 A_1/\hat{\b}^2)$ of the fermion and scalar fields, respectively.

Note that for $A_1=0$ in (\ref{topcurr0}) one has the static version of an important feature of the ATM model, i.e. the classical equivalence between the $U(1)$ Noether $J^{\mu}$ and topological currents $j^{\mu}_{top}$ \cite{matter, npb2}
\br
\label{equi11}
 {\bar{\psi}}\, \gamma^{\mu}\, \psi &=&  \frac{1}{\hat{\b}} \epsilon^{\mu\nu} \pa_{\nu} \, \vp,
\er
with
\br
J^{\mu} &=& {\bar{\psi}}\, \gamma^{\mu}\, \psi,\\
j_{top}^{\mu} &=& \epsilon^{\mu\nu} \pa_{\nu} \, \vp.
\er 
So, for $A_1\neq 0$ the equation (\ref{topcurr0}) represents a deformation of the static version of the currents equivalence (\ref{equi11}) by incorporating a non-local term depending on the scalar potential.

\section{Fermion-kink configurations and spinor zero-modes}
\label{sec:bs}

In order to solve the system of equations (\ref{sta1})-(\ref{sta2}) and (\ref{topcurr0})  we will use the Hirota tau function approach in which the scalar and the spinor components are parametrized by the tau functions as
\br
\label{tau1f}
e^{i \hat{\b} \vp} &=& e^{i \frac{\theta_1}{2}}  \, \,\frac{\tau_0}{\tau_1},\,\,\,\,\, \theta_1 \in \IR, \\
\( \begin{array}{c}
\xi_R \\
\xi_{L} \end{array}\) &=& \sqrt{\frac{m_1}{4}} \(\begin{array}{c}
 \tau_R/ \tau_0\\
- \tau_L/\tau_1 \end{array}\),\,\,\,\,\, \( \begin{array}{c}
\xi^\star_R \\
\xi^\star_{L} \end{array}\) = - \sqrt{\frac{m_2}{4}} \(\begin{array}{c}
 \widetilde{\tau}_R/ \tau_1\\
\widetilde{\tau}_L/\tau_0 \end{array}\),\label{tau1ff}
\er
with $m_1, m_2$ real parameters. Note that the scalar field $\vp$ is guaranteed to be real by taking 
\br
\label{unita1}
\tau_1  = (\tau_0)^{\star}.
\er 
 
Substituting the above parametrization into  (\ref{xis11})-(\ref{xis12})  one gets
\br
\label{tau11}
i \epsilon \tau_R\tau_0 + \tau_0 \frac{d}{dx} \tau_R - \tau_R \frac{d}{dx} \tau_0 &=& e^{-i \theta_1} M \tau_1 \tau_L\\
i \epsilon \tau_L\tau_1 - \tau_1 \frac{d}{dx} \tau_L + \tau_L \frac{d}{dx} \tau_1 &=& - e^{i \theta_1} M \tau_0 \tau_R. \label{tau12}
\er
Similarly, substituting into (\ref{topcurr0}) one gets
\br
\nonumber
&&\frac{- \sqrt{m_1 m_2} \hat{\b}^2 (\widetilde{\tau}_R \tau_R - \widetilde{\tau}_L \tau_L ) + 4i (\tau_0 \pa_x \tau_1 - \tau_1 \pa_x \tau_0)}{4 \hat{\b}^2 \tau_0\tau_1}  -\\
&& (\frac{A_1}{i}) \int_{-\infty}^{x} dx'[  \frac{e^{i\theta_1}\tau_0^4 - e^{-i\theta_1}\tau_1^4}{\tau_0^2\tau_1^2} ] = 0
\label{tau21}
\er

\subsection{Kinks and zero-mode fermion bound states}
\label{subsec:1kinkbs}

Let us assume the following expressions for the tau functions for $1-$kink and the spinor bound states
\br
\label{tau1}
\tau_1 &=& 1 + i e^{2 K x};\,\,\, \tau_0 = 1 -i  e^{2 K x};\\
\label{tau2}
\tau_R &=&   \zeta a_{+}  \,e^{K x } ,\,\,\, \widetilde{\tau}_R = - a_{-}\, e^{K x} ,\\
\label{tau3}
\tau_L &=& - a_{+} \, e^{K x} ,\,\,\, \widetilde{\tau}_L = \, \frac{1}{\zeta} a_{-} \, e^{K x},
\er
with $a^{\star}_{+} = \zeta \sqrt{\frac{m_2}{m_1}} \, a_{-}$. These expressions solve the system of equations  (\ref{tau11})-(\ref{tau12})  and   (\ref{tau21}) provided that the parameters satisfy the next relationships
\br
\label{params}
\epsilon &=& 0,\,\,\,\,\,\,\,\,\,\,\,\, K = \pm  M,\,\,\,\,\zeta = \mbox{sign}[\frac{K}{M}],\,\,\,\,\,\theta_1 = 2\pi n, n \in \IZ,\\
A_1 & = & \frac{M^2}{\hat{\b}} + \frac{1}{8} M a_{+} a_{-} \sqrt{m_1 m_2}.   
\er
The above solution represents a zero-mode $\epsilon =0$. 
 
Notice that taking into account (\ref{unita1}), from (\ref{tau1f}) one can write 
\br
\label{vptau}
\vp =  \frac{2}{\hat{\b}} \arctan{\big\{-i \Big[\frac{  \tau_0 - \tau_1}{  \tau_0 + \tau_1}\Big]\big\}}.
\er
Next, we construct the kink solutions for the tau functions in (\ref{tau1}). One has
\br
\label{vptau1}
\vp^{\pm}(x) = -\frac{2}{\hat{\b}} \arctan{\Big[ e^{2 K x}\Big]},\,\,\,\, K \equiv \pm M. 
\er
This is a kink(anti-kink) solution of the model. These solutions exhibit the topological charges
\br
\label{fractop}
Q_{k(\bar{k})} &=& \frac{\hat{\b}}{2\pi} (\vp^{\mp}(+\infty)-\vp^{\mp}(-\infty)),\,\,\,\,\, k=kink, \,\,\,\bar{k}=antikink\\
&=& \pm \frac{1}{2}  . \label{fractop1}
\er 
Therefore one can consider the state with topological charge $Q_{kink-top}^{(I)} = + \frac{1}{2}$ as the kink and the state with $Q_{kink-top}^{(I)} = - \frac{1}{2}$ as the anti-kink. Note that these charges are fractional, i.e. one-half of the integer $\pm 1$. 
The zero-mode spinor components become
\br
\label{0mode}
\( \begin{array}{c}
\xi_R \\
\xi_{L} \end{array}\) &=&\frac{\sqrt{m_1} a^+}{2} \(\begin{array}{c}
 \frac{\zeta\, e^{\g x}}{1-i e^{2 \g x}} \\
-  \frac{e^{\g x}}{1+i e^{2 \g x}} \end{array}\),\,\,\,\,\ \g = - \mbox{sign} (\zeta) M.
\er

For later purposes, let us define the mass parameter $m$ of the scalar particle such that 
\br
A_1 \equiv \frac{m^2}{4 \hat{\b}^2}.
\er  

In addition, for the kink-type  solution (\ref{vptau1}) one can write the important relationship 
\br
\label{dvp}
\frac{d}{dx} \vp = \frac{2 K}{\hat{\b}} \sin{(\hat{\b} \vp)},\,\,\, K = \pm M.
\er
Using this identity the integro-differential equation (\ref{topcurr0}), in the kink sector, turns out to be 
\br
\label{curr12sin}
\xi_{R}^\star \xi_R + \xi_{L}^\star \xi_L + 2(\frac{K}{\hat{\b}^2}-\frac{A_1}{K}) \sin{(\hat{\b} \vp)} =0.
\er
This is the currents equivalence relationship modified by the presence of the scalar mass parameter term $A_1$. So, using the definition $J^0 = \xi_{R}^\star \xi_R + \xi_{L}^\star \xi_L$ and the relation (\ref{dvp}) into (\ref{curr12sin}) one can write    
\br
\label{J0vp}
J^0 = - \frac{1}{\hat{\b}}(1-\frac{m^2}{4 K^2}) \vp'. 
\er 
Actually, this equation shows the modification of the topological and Noether currents equivalence by the term containing the parameter $A_1$ due to the self-coupling scalar potential in the Lagrangian (\ref{atm0}). In addition, this relationship shows that the ATM modified model (\ref{atm0}) inherits from the ATM model a soliton-fermion duality symmetry \cite{npb1}.   

Let us emphasize that the tau-function formalism has been useful in determining the zero-mode bound state for the spinor coupled to the kink (\ref{vptau1}). However, bound states with $\epsilon \neq 0$ will be examined below in the context of the related Heun's type equations for the spinor components.   

\section{Energies of kink-fermion configurations and spinor bound states}
\label{sec:energy}

In this section we compute the energy of the soliton-fermion configurations in the zero-mode fermion bound state $\epsilon=0$ sector associated to the ATM model (\ref{atm0}). We perform this computation firstly by writing the energy density associated to the Lagrangian (\ref{atm0}) for static configurations, and then specializing the result for the on-shell first order system of equation (\ref{sta1})-(\ref{sta2}). So, from (\ref{atm0}) one can define
\br
{\cal H} = \dot{\vp} \Pi_{\vp} + \dot{\psi}_R \Pi_{R} +  \dot{\psi}_L \Pi_{L}  - {\cal L},
\er
with
\br
\Pi_{\vp} \equiv  \dot{\vp},\,\,\, \Pi_{R} \equiv -i \psi^\star_R,\, \,\,\,  \Pi_{L} \equiv -i \psi^\star_L.
\er
Next, taking into account the Ansatz (\ref{spbs1}) the energy of static configurations (set $\Pi_{\vp} = 0$)  can be written as
\br
E  &=&  \int \, dx \Big\{  \frac{1}{2} \vp'^2 - \pi_{R}  [\pa_x \xi_R + M  e^{i 2\hat{\b} \vp} \xi_L] + \pi_{L}  [\pa_x \xi_L + M e^{-2i \hat{\b} \theta} \xi_R] + A_1 (1-\cos{(2 \hat{\b} \vp)})\Big\}, \label{en0}
\er
where $\pi_R = -i \xi^\star_R$ and $\pi_L = -i \xi^\star_L$.  
In order to compute $E$ we  assume that the static field configurations satisfy the first-order equations (\ref{sta1})-(\ref{sta2}). So, the energy (\ref{en0}) of the kink-fermion static configurations becomes  
\br
\label{en1}
E =  \int \, dx \Big\{  \frac{1}{2} \vp'^2 + \epsilon J^0  + 2 A_1 [\sin{(\hat{\b} \vp)}]^2\Big\},  
\er  
where the potential term has been rewritten for later convenience. Notice that the expression (\ref{en1}) defines the energy of an arbitrary spinor-kink configuration of the model. In particular, for kinks of the sine-Gordon type (\ref{vptau1}) and taking into account its relevant relationship (\ref{dvp}), as well as considering the fermion charge normalization $\int_{-\infty}^{+\infty} dx J^0 =1$, the expression (\ref{en1}) can be written as
\br
\label{en11}
E &=&  \int \, dx  (\frac{1}{2} + \frac{m^2}{8 M^2} )\vp'^2 + \epsilon ,\\
  &=& \int^{\vp(+\infty)}_{\vp(-\infty)} \, d\vp \,  (\frac{1}{2} +\frac{m^2}{8 M^2} ) \frac{2 M}{\hat{\b}} \sin{(\hat{\b} \vp)} + \epsilon \label{en12}\\
&=& E_{kf} + \epsilon \label{en3},\er
with
\br
E_{kf}&\equiv& - \frac{2 \pi M}{\hat{\b}^3} (1+\frac{m^2}{4 M^2} ) Q_{k(\bar{k})}. \label{ekf}
\er
Notice that (\ref{ekf}) follows from the first term in (\ref{en12}) taking into account the Eqs. (\ref{dvp}) and (\ref{fractop}). So, the Eq. (\ref{ekf}) defines the kink-fermion configuration energy. The spinor-kink sector in (\ref{en3}) $E_{kf}$ shows that the energy depends on the topological charge $Q_{k(\bar{k})}$ associated to the scalar kink(antikink) solution. So, in order to compute the energy $E$ in (\ref{en3}) of the whole soliton-spinor configuration, it suffices to know the parameters $Q_{k(\bar{k})}, \hat{\b}, M, m$ and the spinor bound state energy $\epsilon$. 

The explicit spinor contribution $\epsilon J^0$ in (\ref{en1}) vanishes for the zero-mode case ($\epsilon =0$). So, the quantity (\ref{en3}) with $\epsilon=0$ represents the energy of the scalar plus spinor configuration with zero-mode. Since the model (\ref{atm0}) inherits from the ATM model \cite{npb1} a soliton-fermion duality symmetry  (see Eq. (\ref{J0vp})) the energy $E$ can also be written as an integral of an energy density proportional to the squared fermion current $(J^0)^2$ in (\ref{en11}). In fact, one can write $\vp' \sim J^0$ due to the currents equivalence (\ref{J0vp}), and then the  calculation performed in this way provides the same result as (\ref{en3}). 

Let us discuss the stability of the above spinor-fermion solution in the zero mode sector $\epsilon =0$. So, examining the solution to the equation $E'(M)=0$ one finds that 
\br
\label{Mm2}
\frac{dE}{dM}|_{M=M_o}=0  \rightarrow M_o = \frac{m}{2}
\er
At this stage, the spinor–kink configuration attains stability at $m = 2 M_o$. Interestingly, the very presence of this point relies on a nonvanishing mass parameter $m$ for the scalar boson. Nevertheless, as we demonstrate below, a proper assessment of how fermionic backreaction affects the stability and dynamics of fermion–kink systems requires incorporating both the valence fermion energy and the fermion vacuum polarization energy (VPE). Moreover, the requirement $ m \neq 0$ is physically consistent, since a massless scalar field in $1+1$ dimensions would lead to infrared divergences. The self-interaction term therefore plays a dual role, ensuring both soliton stability and infrared regularity.

A similar relation between the scalar and fermion masses, Eq. (\ref{Mm2}), was also reported in a recent study \cite{loginov}. In that work, however, fermion scattering on a sine-Gordon kink was treated within the external-field approximation, meaning that the soliton background was prescribed a priori. In contrast, in our model the fermion–soliton configuration, Eqs. (\ref{vptau1}) and (\ref{0mode}), constitutes an exact solution that fully incorporates the spinor backreaction on the soliton. In the approach of \cite{loginov}, the emergence of the relation (\ref{Mm2}) followed from truncating the series expansion of the Heun function describing the spinor wave, a step necessary to obtain a zero-mode bound state. We elaborate on this point-and on the appearance of an additional valence fermion-in the next section.
    
Let us emphasize that the system of first-order equations (\ref{sta1})–(\ref{sta2}) plays a role analogous to that of the Bogomolny-Prasad-Sommerfield (BPS) equations. Specifically, they not only reproduce the second-order Euler–Lagrange equation governing the scalar field, but also fix the total energy (\ref{en3}) in terms of the associated topological charges. These first-order relations are intrinsically linked to the structure of the energy functional (\ref{en0}) and the corresponding static energy (\ref{en1}). Within that framework, the BPS bounds provide a particularly effective method for constructing topological soliton solutions, as they constrain the soliton energy by topological considerations. Configurations that saturate this bound necessarily obey a set of first-order differential equations, namely the BPS equations.

\section{Scattering and bound states via the Heun-equation approach}

\label{sec:Heun}

The presence of the soliton distorts the fermionic mode structure, leading to a spectrum that includes soliton-induced bound states and altered scattering states absent in the free theory. This interaction energy between the kink and the Dirac vacuum is a crucial component for maintaining the internal consistency of the fermionic semiclassical expansion. Below we will compute the scattering states of the fermion-soliton configuration of the modified ATM model (\ref{atm0}). 

We will concentrate on the spinor scattering states with backgroud soliton field (\ref{vptau1}). So, let us consider the two-component spinor parameterized as
\br
\label{spbs11}
\psi   = e^{- i E_1 t}\(\begin{array}c
u(x)\\
v(x)\end{array}\),
\er
where the spinor components $u$ and $v$ define the scattering solutions in the presence of the soliton $\vp$, and $E_1$ is the energy of these states defined as $E_1^2 = M^2 + k^2$. 
 
So, from (\ref{xis11})-(\ref{xis12}) and (\ref{curr12sin})-(\ref{J0vp}) one can write the coupled system of static equations
\br
\label{sta11}
- E_1 u + i \pa_x u +i  M e^{-2i \hat{\b} \vp} v &=&0,\\
\label{sta21}
E_1 v + i\pa_x v +i M e^{2i \hat{\b} \vp} u &=&0,\\
\Big\{u^\star u + v^* v - \Big[u^{\star\,(free)}  u^{\,(free)}  + v^{\star\,(free)} v^{(free)} \Big](x = +\infty)\Big\} + \nonumber \\
\frac{1}{\hat{\b}}(1-\frac{m^2}{ 4 K^2}) \vp'&=& 0, \label{sta31}
\er  
where the symbol $^{\star }$ stands for complex conjugation as usual. Notice that in (\ref{sta31}) it has been subtracted the contribution of 
the constant charge density due to the free state $(u^{(free)}, v^{(free)}  )^T$ evaluated at $x= + \infty$, such that the equation  becomes consistent in the asymptotic regions of the soliton at $+ \infty$. A similar boundary condition was recently employed in \cite{prd2} for a model in which the solitons possess varying topological charge. 

Next, let us obtain a second order differential equation associated to the $u$-component. So, from (\ref{sta11}) one can write
\br
\label{vx}
v(x) = -\frac{i}{M}\,  e^{2i \hat{\b} \vp(x)} \, (E_1 u - i u'(x)).
\er 

We will assume the following asymptotic forms
\br
\left(\begin{array}{c}
u(x) \\
v(x)
\end{array}\right) &\xrightarrow[x \rightarrow -\infty]{\,}& \left(\begin{array}{c}
 c_1 \, e^{i k x} +   c_2 \,e^{\frac{\pi k}{2M}}\, e^{-i k x} \\
  d_1 \, e^{i k x}  +  \,d_2\, e^{-i k x} 
\end{array}\right),\label{left1}\\
\left(\begin{array}{c}
u(x) \\
v(x)
\end{array}\right) &\xrightarrow[x \rightarrow +\infty]{\,}& \left(\begin{array}{c}
  c_o e^{i k x}\\
   d_o e^{i k x}  
\end{array}\right), \label{right1} 
\er
Notice that at $x= + \infty$ one has the transmitted wave components  and at $x= - \infty$  the both incident and reflected components. Unitarity requires the coefficients of the transmitted wave to satisfy  $|c_o|^2 + |d_o|^2 =1$.

Substituting the relationship for $v$ (\ref{vx}) into (\ref{sta21}) and inserting the sine-Gordon type soliton (\ref{vptau1}) into the scalar field, one gets 
\br
\label{u1x}
u''(x) - 4i K \mbox{sech}(2 K x) u'(x) + (E_1^2-M^2 +4 E_1 K \mbox{sech}(2 K x)) u(x) =0.
\er
The solutions to the differential equation (\ref{u1x}) can be 
expressed in terms of the local Heun functions \cite{ronveaux, olver, slavyanov}. This equation has recently been considered in the study of an external sine-Gordon soliton coupled to a spin-isospin fermion model \cite{loginov}. After the relevant parameter identifications, the system of equations (\ref{sta11})-(\ref{sta21}) and the corresponding second order differential equation (\ref{u1x}), coincide with the model considered in \cite{loginov} for the diagonal spinor components $(u, v)$ of a $2\times 2$ spin-isospin matrix. Our approach, however, differs in that the soliton–fermion configuration is handled analytically and exactly, meaning that the soliton profile is not prescribed a priori. Our exact treatment shows that the soliton width, proportional to 
$\frac{1}{2 K}$, depends on the fermion mass. In contrast, in the external-field approximation used in \cite{loginov}, the width is assumed to be determined by the scalar particle mass $m$ as $\sim \frac{1}{m}$.

\subsection{Scattering states and the Heun's local functions}

The equation (\ref{u1x}), after successive transformations, can be written as the next general Heun equation
\br
H''(z) + P(z) H'(z)+ Q(z) H(z) &=&0, \label{heun0}\\
\label{PP1}
P(z) &=& (\frac{\g}{z}+ \frac{\d}{z-1}+\frac{\epsilon}{z-a})\\
Q(z) &=&  \frac{\a \b z - q}{z(z-1)(z-a)},\label{QQ1}
\er
with  
\br
\label{abgd1}
\g + \d + \epsilon = \a + \b +1.
\er
The local Heun function $Hl[a, q; \a, \b, \g, \d;z]$ \cite{ronveaux, slavyanov} is defined as the Frobenius solution of the general Heun equation (\ref{heun0}) regular at $z=0$. The differential equation (\ref{heun0}) possesses four regular singular points at $z=0,1,a,\infty$. Consequently, the radius of convergence of the local series around $z=0$ is determined by the distance to the nearest singular point and equals $\min{\{1, |a|\}}$. Thus the parameter $a$, which fixes the position of the fourth singularity, controls the convergence domain of the series representation.

The specific parameter sets will be specified below according to the  incident/reflected and transmitted scattering channels.

So, the problem under consideration constitutes a well-posed scattering formulation for the general Heun equation, and then the solutions to the differential equation (\ref{u1x}) can be expressed in terms of local Heun functions \cite{ronveaux, slavyanov}. The analysis proceeds in three stages:
(i) identification of the parameters governing the incident, reflected and transmitted states in the asymptotic regimes $ x \rightarrow \pm \infty$. 
(ii) determination of the local Frobenius exponents and characterization of the two linearly independent local solutions at the regular singular point $z=0$, corresponding to the scattering asymptotic regions $|x| \rightarrow \infty$. At a regular singular point the solutions of (\ref{heun0}) behave as power laws and admit convergent Frobenius series. At the point $z=z_0=0$ one has that $(z-z_0) P(z) = \g + {\cal O}(z)$ and $(z-z_0)^2 Q(z) = \frac{q}{a} z +  {\cal O}(z^2)$ and, so, the point $z=0$ is dubbed as a regular singular point ($P, Q$ are defined in (\ref{PP1})-(\ref{QQ1})). And
(iii) examination of the resulting scattering data, including the asymptotic behavior, the extraction of reflection and transmission coefficients, and the conditions for the existence of normalizable (bound) states.
 
For a fermionic mode incident on the kink from the left for $K < 0$, let us consider the solution of Eq. (\ref{u1x}) in terms of the Heun's local function, such that asymptotically it must reduce to the transmitted plane wave, being  $\sim e^{ik x}$. Then, the corresponding transmitted wave becomes
\br
\label{ur10}
u_{tr}(x) = \, e^{ik x}\, Hl[\frac{1}{2}, -i \frac{E_1+k}{K}; -1, 0, 1-i\frac{k}{K}, 1+i\frac{k}{K}; \frac{1}{1+i e^{-2 K x}}],\label{ur1}
\er
where the notation $Hl[a=\frac{1}{2}, q; \a, \b, \g,\d; z]$ is used for the six-parameter local Heun function \cite{ronveaux, slavyanov}. The behavior of $z$ in the original spatial coordinate becomes $z \rightarrow -i e^{2 K x}$ ($K<0$) in the limit $ x \rightarrow + \infty$. So, the argument of the local Heun function tends to zero (one of the regular singular points at $z=0$) as $ x \rightarrow +\infty$. The local Heun function 
$Hl[\frac{1}{2}, q; \a, \b, \g,\d; z]$  is analytic at the regular singular point $z=0$ and it takes the constant value 
$Hl(0)= c_0$ (below we will define $c_0 =1$ for the scattering states). Around this point, it admits a convergent Taylor expansion. In the complex $z-$plane the radius of convergence of this local series is given by $min\{|a|, 1\}$; for $a=1/2$, this yields a convergence radius of $1/2$.

Although the Taylor expansion of 
$Hl[a, q; \a, \b, \g,\d; z]$ possesses only a finite radius of convergence, the function itself admits analytic continuation to the entire complex plane, with a branch cut conventionally taken along 
$[\frac{1}{2}, \infty]$. It is thus well-defined at all finite points of the complex plane except at the regular singularities $z=\frac{1}{2}$ and $z=1$. Note that in (\ref{ur10}), the argument of the local Heun function approaches unity as $x\rightarrow -\infty$ for $K<0$, implying that these local solutions cease to provide a valid representation in this asymptotic regime.

However, there are two linearly independent solutions with analytic behavior at $x \rightarrow  -\infty$. They become
\br
u_{in}(x) &=&   e^{i k x} Hl[\frac{1}{2}, i \frac{E_1 +k}{K}; -1, 0, 1-i\frac{k}{K}, 1+i\frac{k}{K}; \frac{1}{1-i e^{2 K x}}]\label{uright12}
\er
and
\br
u_{ref}(x) &=&\nonumber  e^{- i k x} (-i+ e^{-2 K x})^{-\frac{i k}{K}} \times\\
&& Hl[\frac{1}{2}, i\frac{2E_1-k}{2 K} - \frac{k^2}{2 K^2}; -1+i\frac{k}{K}, i\frac{k}{K}, 1+i\frac{k}{K}, 1+i \frac{k}{K}; \frac{1}{1-i e^{2K x}}]. \label{ul12}
\er 
Note that the argument of the local Heun functions above tend to zero (regular singular point at $z=0$) as $ x \rightarrow -\infty$.

Then, one has the incoming wave $u_{in}(x)$ in (\ref{uright12}) and the reflected wave $u_{ref}(x)$ in (\ref{ul12}). The outgoing wave $u_{tr}(x)$ has been presented in the Eq. (\ref{ur1}). 
  
Therefore, the general local solution around the regular singular point $z=0$, i.e. at the regions $x \rightarrow -L$ ($L>>0$) for $K<0$, will be a linear combination of (\ref{uright12}) and (\ref{ul12})
\br
\label{u12}
u_s(x) = c_1 u_{in}(x) + c_2 u_{ref}(x). 
\er  
Moreover, the solutions given in Eqs. (\ref{ur1}), $u_{tr}$, and $u_s$ in (\ref{u12}) possess overlapping domains of analyticity in the variable $x$
. By equating these representations $u_{tr}|_{x=x_0}=u_s|_{x=x_0}$ 
together with their first derivatives $u'_{tr}|_{x=x_0}=u'_s|_{x=x_0}$ at the point $x=x_0$, one can determine the coefficients $c_1$ and $c_2$ appearing in Eq. (\ref{u12}). The specific choice of the matching point is immaterial, provided it lies within the interval $(-\infty, + \infty)$; for convenience and by symmetry, we select $x_0=0$ as the matching point. So, one has \cite{loginov, chai, chen}
\br
\label{match1}
u_{tr}|_{(x=0)} &=& c_1 u_{in}|_{(x=0)} + c_2 u_{ref}|_{(x=0)},\\
\frac{d}{dx}u_{tr}|_{(x=0)} &=& c_1 \frac{d}{dx} u_{in}|_{(x=0)} + c_2 \frac{d}{dx} u_{ref}|_{(x=0)}.\label{match2}
\er
This system provides the explicit expressions
\br
\label{match11}
c_1 = \frac{W(u_{tr}, u_{ref})|_{x=0}}{W(u_{in}, u_{ref})|_{x=0}},\,\,\,\,\,\,\,\, c_2 = -\frac{W(u_{tr}, u_{in})|_{x=0}}{W(u_{in}, u_{ref})|_{x=0}},
\er
where $W(v,w) \equiv v w'-w v'$ is the Wronskian of the two functions $v$ and $w$. 

The Wronskians $W(u_{tr}, u_{in})|_{x=0}$, $W(u_{tr}, u_{ref})|_{x=0} $ and $W(u_{in}, u_{ref})|_{x=0} $ become
\br
\label{wr1}
W(u_{tr}, u_{in})|_{x=0}  &=& M [h_{tr}(s) h'_{in}(s^*) + h'_{tr}(s) h_{in}(s^*)],\,\,\,\,\,\,\,\,\,\,\,\,\,\,\,\, s = \frac{1-i}{2},\\
\label{wr2}
W(u_{tr}, u_{ref})|_{x=0}  &=& 2^{\frac{i k}{2K}} e^{-\frac{\pi k}{K}} [ 2 k s^* h_{tr}(s) h_{ref}(s^*) + K (h_{tr}(s) h'_{ref}(s^*)+h'_{tr}(s) h_{ref}(s^*) ) ],\\
W(u_{in}, u_{ref})|_{x=0}  &=& 2^{\frac{i k}{2K}} e^{-\frac{\pi k}{K}} [ 2 k s^* h_{in}(s^*) h_{ref}(s^*) + K (h_{in}(s^*) h'_{ref}(s^*)- h'_{in}(s^*) h_{ref}(s^*)) ], \label{wr3}
\er
where
\br
\label{hn1}
h_{tr}(s)&=&Hl[\frac{1}{2}, -i \frac{E_1+k}{K}; -1, 0, 1-i\frac{k}{K}, 1+i\frac{k}{K}; s],\\
\label{hn2}
h_{in}(s^*) &=& Hl[\frac{1}{2}, i \frac{E_1 +k}{K}; -1, 0, 1-i\frac{k}{K}, 1+i\frac{k}{K}; s^*],\\
h_{ref}(s^*) &=& Hl[\frac{1}{2}, i\frac{2E_1-k}{2 K}- \frac{k^2}{2 K^2}; -1+\frac{ik}{K}, \frac{i k}{K}, 1+\frac{ik}{K}, 1+\frac{i k}{K}; s^*] . \label{hn3}
\er
Therefore, from  (\ref{match11}) and taking into account (\ref{wr1})-(\ref{wr3}) and (\ref{hn1})-(\ref{hn3}) one has
\br
\label{c111}
c_1(k) &=& \frac{2k s^* h_{tr}(s) h_{ref}(s^*) + K (h'_{tr}(s) h_{ref}(s^*)+h_{tr}(s) h'_{ref}(s^*))}{2k s^* h_{ref}(s^*) h_{in}(s^*)- K (h_{ref}(s^*) h'_{in}(s^*)-h'_{ref}(s^*) h_{in}(s^*))},\\
c_2(k) &=& e^{-\frac{\pi k}{4 K}} 2^{i \frac{k}{2K}} \Big[\frac{- K (h'_{tr}(s) h_{in}(s^*)+h_{tr}(s) h'_{in}(s^*))}{2k s^* h_{ref}(s^*) h_{in}(s^*)- K (h_{ref}(s^*) h'_{in}(s^*)-h'_{ref}(s^*) h_{in}(s^*))}\Big],
\label{c222}
\er
where the expressions  for $h_{tr}(s), h_{ref}(s^*), \mbox{and} \, h_{in}(s^*)$ are provided in (\ref{hn1})-(\ref{hn3}).

The coefficients 
$c1(k)$ and $c2(k)$ encode the full scattering data associated with the interaction of the fermionic wave with the sine-Gordon–type kink. In Fig. 1, we display the real and imaginary components of the scattering wave function $Re(u)$ and $Im(u)$, together with the kink configuration of topological charge $Q_{k(\bar{k})} = \frac{1}{2}$
. The figure provides a qualitative illustration of the influence of the matching conditions (\ref{match1})–(\ref{match2}) on the behavior of the scattering wavefunction $u$ at the origin $x=0$.

\begin{figure}
\centering
\label{fig1}
\includegraphics[width=1.5cm,scale=4, angle=0,height=4.5cm]{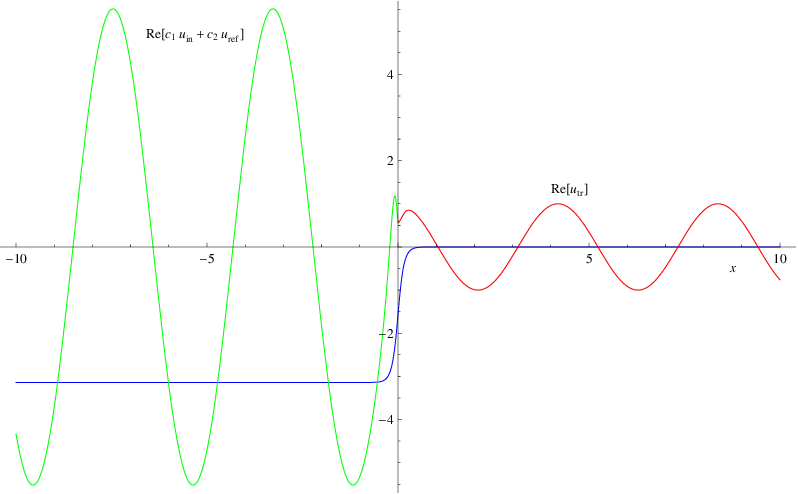}
\includegraphics[width=1.5cm,scale=4, angle=0,height=4.5cm]{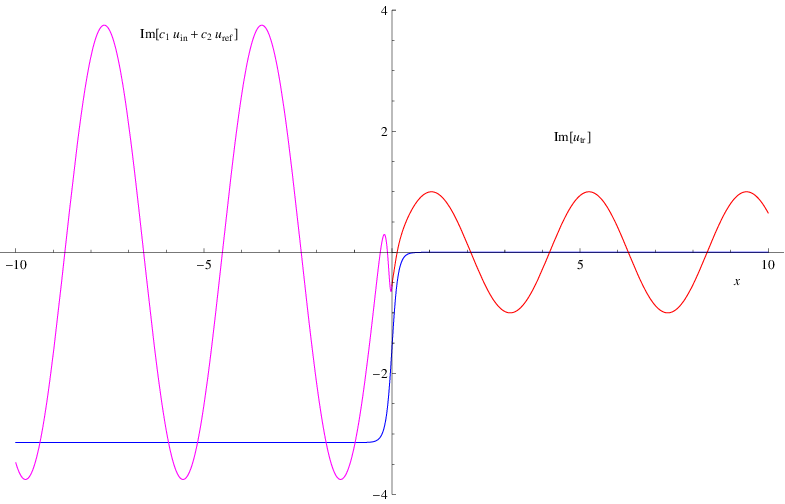}  
\parbox{6in}{\caption{(color online) Scalar kink (blue) in the left and right Figs. The real and imaginary parts of the scattering wave function $u(x)$: Left Fig. $Re[c_1 u_{in}(x) +c_2 u_{ref}(x)]$ (green), $Re[u_{tr}(x)]$ (red), and  right Fig. $Im[c_1 u_{in}(x) +c_2 u_{ref}(x)]$ (magenta), $Im[u_{tr}(x)]$ (red). For $\b =1, K=-5, k=1.5, E_1=+5.22$.}}
\end{figure}

From Eqs. (\ref{ur1}) and (\ref{u12}), it follows that the asymptotic behavior of the $u-$component of the spinor wave function can be schematically expressed as (after multiplied by an overall factor $\frac{1}{c_1}$)
\br
\label{asympt1}
 e^{ikx} \rightarrow  \frac{1}{c_1} e^{ikx} + \frac{c_2}{c_1} e^{\frac{\pi k}{2 M}}e^{-ikx},
\er
which describes the decomposition of the incident wave into its transmitted and reflected components.
  
During the scattering process, the transmitted fermionic wave component $u$ acquires a phase shift $\d_{u}$ relative to the incident wave.  According to Eq. (\ref{asympt1}), this phase shift is given by
\br
\label{phs1}
\delta_u(k) = -\arg{[c_1(k)]}.
\er
Similarly, from Eqs. (\ref{ur1}) and (\ref{u12}), and taking into account (\ref{vx}) the asymptotic behavior of the $v-$component of the spinor wave function can be schematically expressed as
\br
\label{asympt2}
c_1   e^{ikx} \rightarrow   e^{ikx} + c_2  (\frac{E_1- k}{E_1+k})\, e^{\frac{\pi k}{2 M}}e^{-ikx},
\er
such that the incident wave is decomposed into its transmitted and reflected components.

Then, the transmitted fermionic wave component $v$ acquires a phase shift $\d_{v}$ relative to the incident wave. According to Eq. (\ref{asympt2}), this phase shift is given by
\br
\label{phs2}
\delta_v(k) = -\arg{[c_1(k)]}.
\er 
One can conclude that the upper and lower spinor components develop the same phase shift after the scattering on the sine-Gordon type soliton. This is in contradistinction to the result recently obtained in the ATM model with variable topological charge soliton background, in which the phase shifts of the $u$ and $v$ components differ by a constant. Actually, in that case a unique phase shift has been defined as the average of the upper and lower component phase shifts \cite{prd2}.   

The scattering states for the lower component $v$ can be obtained directly from the expressions above for the upper component $u$ via Eq. (\ref{vx}). So, one has
\br
\label{vtr1}
 v_{tr}(x) &=&  \frac{i e^{i k x}}{K} e^{2i\b \vp} \Big[(E_1+k)  h_{tr}(z_1) - i K h'_{tr}(z_1) \frac{1}{1-i \sinh{(2K x)}} \Big],\,\,\,\,\,\,\,\,\,\,\,\, z_1(x) \equiv \frac{1}{1+i e^{-2K x}},\\
\label{vin1}
v_{in}(x) &=&   \frac{i e^{i k x}}{K} e^{2i\b \vp} \Big[(E_1+k)  h_{in}(z_2) + i K h'_{in}(z_2) \frac{\mbox{sech}^2( K x)}{(1-i \tanh{(K x)})^2} \Big],\,\,\,\,\, z_2(x) \equiv \frac{1}{1-i e^{2 K x}},\\
\nonumber
v_{ref}(x) &=&   \frac{i e^{-i k x} e^{2i\b \vp}}{K} (-i + e^{-2 K x})^{-\frac{i k}{K}} \times\\
 &&\Big[ (E_1 -i k \frac{\tanh{K x} -i }{\tanh{K x} +i})  h_{ref}(z_2) +i K h'_{ref}(z_2) \frac{\mbox{sech}^2(K x)}{(1-i \tanh{(K x)})^2} \Big].\label{vrf1}
\er  

The above scattering states for the spinor components must satisfy (\ref{sta31}). Thus, for the transmitted component one has
\br
u^{\star}_{tr}(x) u_{tr}(x) + v^{\star}_{tr}(x) v_{tr}(x) -[1 + (\frac{E_1 + k}{K})^2] = - \frac{1}{\hat{\b}}(1-\frac{m^2}{4 K^2}) \vp'(x),\label{trcons}
\er 
where the boundary condition $[(u^{free}_{tr})^\star u_{tr}^{free} + (v^{free}_{tr})^{\star} v^{free}_{tr}]_{(x\rightarrow +\infty)} = 1 + (\frac{E_1 + k}{K})^2$ has been used.

Then, in order to find an additional relationship between the parameters, one can use the expression (\ref{trcons}) evaluated at $x=0\,[s = z_1(x=0)]$ for simplicity. So, one has 
\br
[1 + (\frac{E_1+k}{K})^2 ]|h_{tr}(s)|^2 + |h'_{tr}(s)|^2 &+& 2(\frac{E_1+k}{K}) Im[h'_{tr}(s) h^{\star}_{tr}(s)] -\\
&&[1 + (\frac{E_1 + k}{K})^2] = \frac{2 K}{\hat{\b}^2} (1- \frac{m^2}{4 K^2}),\,\,\, s = \frac{1-i}{2}.
\er

Next, the fermionic current in the l.h.s. of (\ref{trcons}) can be computed for the incident, transmitted, and reﬂected 
current components. So, taking into account the expressions (\ref{ur10})-(\ref{ul12}) for the upper $u$ component  and (\ref{vtr1})-(\ref{vrf1}) for the lower component $v$, one has
\br
j_{in} &=& |c_1|^2 \(1+ \frac{1}{M^2}(E_1+ k)^2\),\\
j_{ref} &=& |c_2|^2 e^{\frac{\pi k}{M}} \(1+\frac{1}{M^2}(E_1- k)^2\)\\
j_{tr} &=& 1+ \frac{1}{M^2}(E_1+ k)^2.
\er
Then, using the last relationships one can write the transmission and reﬂection coefﬁcients as
\br
T &=& \frac{j_{tr}}{j_{in}} = \frac{1}{|c_1|^2},\\
R &=&  \frac{j_{ref}}{j_{in}} = \frac{|c_2|^2}{|c_1|^2} e^{-\frac{\pi k}{K}} \(\frac{(E_1-k)^2+K^2}{(E_1+k)^2+K^2}\).
\er 
In Fig. 2 we plot theses coefficients for three values of the mass parameter $M$. Notice that these coefficients satisfy $R + T =1.$ It is evident that, at fixed $k$, the transmission coefficient increases as the mass $M$ decreases.

\subsection{Fermionic bound states}

We now consider the fermionic bound states associated with the kink background in the Heun-function and Heun-polynomial approaches. A spectral analysis of Eq. (\ref{u1x}) in the asymptotic regimes $|x| \rightarrow \infty$ reveals that the continuum (scattering) spectrum is characterized by
\br
\label{E11}
E_1^2 = M^2 + k^2, \,\,\,\, k \in \IR, 
\er 
whereas the discrete (bound-state) spectrum satisfies
\br
\label{Ebs}
E_{bs}^2 = M^2 - \kappa^2, \,\,\,\, \kappa > 0. 
\er
In spectral terms, the bound states correspond to a purely imaginary continuation of the quasi-momentum into the complex plane, $k \rightarrow i\kappa$, thus appearing as isolated eigenvalues below the continuum threshold. So, from (\ref{vx}) and (\ref{u1x}) one has that the components $u$ and $v$ of a fermionic bound state behave as $e^{- \kappa |x|} $ as $|x|\rightarrow +\infty $.

Let us analyze the bound states by examining the behavior of the scattering states under the replacement $k \rightarrow i \kappa$.  In Eq. (\ref{ur10}), the argument of the local Heun function approaches zero as $x\rightarrow + \infty$, causing the function to tend to unity. Consequently, under the substitution $k \rightarrow i \kappa$, the transmitted fermionic wave acquires the correct bound-state asymptotic behavior, proportional to $e^{- \kappa x}$ as $x\rightarrow + \infty$. When $k$ is replaced by $i \kappa$ in Eq. (\ref{ul12}), the reflected wave exhibits the correct bound-state asymptotics, $e^{\kappa x}$ as $x\rightarrow - \infty$. The incident wave, however, becomes 
$e^{- \kappa x}$, which diverges in the same limit. To remove this incorrect behavior of the incident fermionic component, the coefficient 
 $c_1(E_1, k)$ must be zero at 
$E_1 = E_{1n}$, $k = i \kappa = i \sqrt{M^2-E_{1n}}$, where $E_{1n}$ is the energy of the bound state. Note that the coefﬁcient $c_1(E_1, k)$ is 
explicitly determined by Eq. (\ref{c111}). Thus, 
the energy levels $E_{1n}$ of the fermionic bound states 
are determined by the solutions of the transcendental equation
\br
\label{c1110}
c_1(E_1, i \sqrt{M^2 - E_1}) =0.
\er

\begin{figure}
\centering
\label{fig2}
\includegraphics[width=1.5cm,scale=4, angle=0,height=4.5cm]{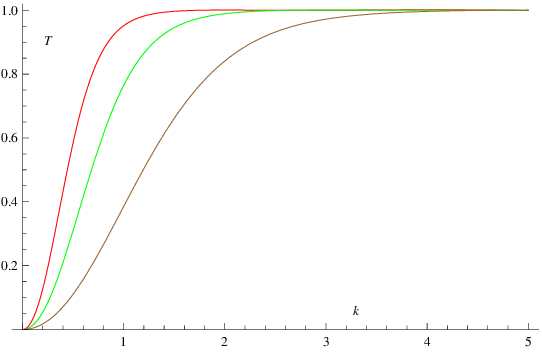}
\includegraphics[width=1.5cm,scale=4, angle=0,height=4.5cm]{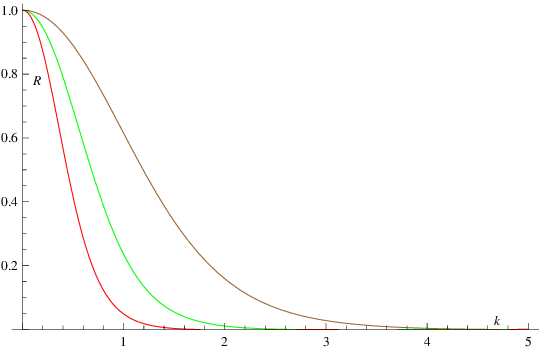}  
\parbox{6in}{\caption{(color online) Dependence of the transmission coefficient $T$ (left Fig.) and the reflection coefficient $R$ (right Fig.) for the fermionic wave function on the fermion momentum $k$ for  the fermion masses $M=0.65$ (red), $M=1$ (green) and $M=1.75$ (brown).}}
\end{figure}
In order to determine the energy eigenvalues we employed symbolic  routines implemented in the Mathematica software package \cite{wolfram}. We consider the Heun functions to be approximated by polynomials of order $N$ in powers of $z$ valid around $z=0$, representing the relevant scattering states. Then, examining the form of the coefficient $c_1$ in the equation (\ref{c111}) one can write in the form 
\br
\label{c1fr}
c_1(E_1, k) =\frac{\Pi_{n=1}^{N} (k - i\, n\, K)}{\Pi_{n=1}^{N} (k + i\, n\, K)}\,\,\, \widetilde{c}_1(k,E_1,K),
\er
where $N$ is the order of the polynomial in the powers of $z$ assumed as approximations to the Heun functions around $z=0$ for the relevant  transmitted and reflected waves, respectively. The factor $\widetilde{c}_1(k,M,K)$ is a rational function of the variables $\{k, E_1, K\}$. In our numerical simulations we choose $N=30$, for which satisfactory convergence is achieved.
 
For $K<0$, the numerator of (\ref{c1fr}) has zeros at 
$ k_n = i\, n\, K$ located on the negative imaginary axis of the complex 
$k$-plane. In contrast to bound states, which correspond to poles of the transmission amplitude, or equivalently zeros of 
$c_1$, on the positive imaginary axis, these zeros signal the presence of virtual states \cite{zavin}. Their wave functions are not square-integrable, since they grow exponentially at infinity, and therefore they do not correspond to physical asymptotic states, although they can substantially affect near-threshold scattering. This interpretation is supported by our numerical results: for  $|K| \leq M$, the phase shift displays sharp variations near  $k=0$, as expected in the presence of virtual states. The Fig. 3 shows the behavior of phase shift $\d(k)$ vs $k$ of the scattering states $u$ and $v$ for  the fermion mass $\,M=1$ and $K= 0.45$. Note that the phase shift shows a marked variation in the low-momentum region.

\begin{figure}
\centering
\label{fig3}
\includegraphics[width=1.5cm,scale=4, angle=0,height=4.5cm]{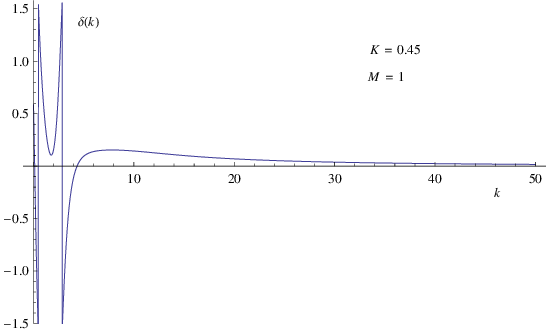}
\parbox{6in}{\caption{(color online) Phase shift $\d(k)$ of the scattering states $u$ and $v$ as a function of $k$, for $\,M=1$ and $K= 0.45$. The phase shift exhibits a sharp variation near $k=0$ in the regime  $|K| \leq M$.}}
\end{figure} 

Therefore, one must search for bound states as the zeroes of the quantity  $\widetilde{c}_1(k,E_1,K)$. Our numerical simulations provide the bound states
\br
\label{bsnumeric}
E_{11} = 0,\,\,\,\,\, E_{12} = \pm 0.986 M.
\er
The first solution $E_{11}$ is the zero-mode bound state. The solutions $E_{12}$ are the non-zero bound-states. Notice that these states correspond to square-integrable wave funcions and genuine asymptotic particles. In this case, the energy values are very close to the threshold fermion energies $E_{thre} \equiv \pm M$.   

A zero mode has already been obtained in the Eq. (\ref{params}) of section \ref{subsec:1kinkbs} for $K= \pm M$, in the framework of the Hirota-tau function formalism. Within numerical accuracy and the relevant parameter identifications, this result coincides with the bound state energies of the diagonal components of the spin–isospin matrix reported in Ref. \cite{loginov}. We note, however, that in our case the soliton width ($\sim \frac{1}{2 M}$) is fixed, whereas in that reference it depends on the scalar mass as $1/m$. Consequently, their model contains an additional free parameter $m$, leading to a one-parameter family of bound states.   
 
The bound-state wave functions associated with the energies in Eq. (\ref{bsnumeric}) have been constructed by imposing continuity conditions that match two locally analytic solutions of Eq. (\ref{u1x}) at a chosen intermediate point. Simultaneously, in the Heun function formalism certain bound states have wave functions that remain analytic over the entire complex plane, excluding only the point at infinity. Below we will show that this is the case for the zero-mode state using the polynomial Heun series solution. 

Moreover, the tau-function result for the zero modes indicates that certain bound states possess wave functions that are analytic throughout the entire complex plane, except at infinity. This situation also arose for the $E_1 \neq 0$ bound-state excitations recently studied in \cite{prd2} for solitons with variable topological charge, where the tau-function method yielded a family of scattering and bound states analytic over the whole complex plane. In the Heun equation  formalism we will see below that the zero-mode emerges as a polynomial Heun series.  

\subsubsection{The zero-mode bound states and polynomial Heun series $Hp_{N}^{(I)}$}

Polynomial (finite-series) solutions of the Heun equation play a crucial role in identifying fermionic bound states in Dirac–soliton systems that are analytic across the entire complex plane. As we will show below, determining the zero-mode bound state of our model is particularly useful, serving as a complementary result to that obtained above through the tau-function approach.

The Heun functions depend on the continuous parameters $K, k, M$, entering the exponent parameters $\a, \b, \g, \d, \epsilon$ and the accessory parameter $q$. In particular, the parameter $q$ depends on the energy $E_1$ of the fermion. However, physical bound states correspond to normalizable spinor solutions, i.e. wavefunctions that decay at spatial infinity. This requirement is satisfied only for specific discrete parameter values, for which the Heun function  truncate to a polynomial as
\br
\label{hpNI}
Hp_{N}^{(I)}[a, q; \a, \b, \g,\d; z] = h_0 + h_1 z + h_2 z^2+...+ h_N z^N,
\er
where
\br
\a = -N,\, (N\in \IN),  
\er
and $h_{N+1}(q)=0$ is a condition fixing the accessory parameter $q$. Therefore, one has the bound state
\br
\label{ubs1}
u_{b}(z) = e^{-\kappa x} \, Hp_{N}^{(I)}(z),
\er
where $Hp_{N}^{(I)}(z)$ is the polynomial Heun function of  type $I$  and degree $N$, according to the classification in Ref. \cite{ronveaux}. 

Since $\a = - 1$ one has $N=1$. Then
\br
\label{hpNI1}
Hp_{1}^{(I)} (z) = h_0 + h_1 z,
\er
where the coefficient parameters $h_0$ and $h_1$, together with the form of $h_{2}(q)$, must be determined. A direct method for doing so is to substitute (\ref{ubs1}) and (\ref{hpNI}) into Eq. (\ref{u1x}). Therefore, one gets the relationship 
\br
\label{det1}
h_{2}(q(E_1)) &\equiv&  E_1 (E_1 - k),\\
  &=&0,\label{det11}
\er
which implies $E_{10} =0$. The coefficients satisfy  
\br   
\label{h10}  
h_1 &=& -\frac{2 \kappa}{M + \kappa} h_0.\,\,\,\,
\er
One can use (\ref{Ebs}) in order to write $ E_{bs} =0 \rightarrow \kappa = M $. Notice that from (\ref{h10}) one can write $h_0+h_1 =0$. Therefore, the polynomial $Hp_{1}^{(I)}(z)$ in (\ref{hpNI1}) vanishes at the point $z=1$. This point corresponds to the variable $x$ in the limit $x\rightarrow -\infty$; so, this is consistent with the vanishing of the bound state $u_{b}(z) $ (\ref{ubs1}) in this limit.

Then, from (\ref{ubs1})-(\ref{hpNI1}) the solution for the zero-mode bound state spinor becomes 
\br
\(\begin{array}{c}
u_{bs}\\
v_{bs}
\end{array} \) = -i u_0 e^{-i \frac{\pi}{4}} 
\(\begin{array}{c}
\frac{e^{M x}}{1-i e^{2 M x}}\\
\frac{e^{M x}}{1+i e^{2 M x}}
\end{array} \),
\er 
where the component $v_{bs}$ has been constructed by substituting $u_{bs}$ into the relationship (\ref{vx}). Remarkably, it is the same zero-mode bound state which has been obtained through the tau-function approach in (\ref{0mode}) provided that  $\zeta=1$ and $u_0 = - \frac{1}{2} \sqrt{m_1} a_+ e^{\frac{3\pi i}{4}}$.

\section{Fermion vacuum polarization energy (VPE)}
\label{sec:vpe}
The vacuum polarization energy (VPE), which includes contributions from infinitely many modes, must be carefully regularized and renormalized to remove divergences. After renormalization, only the finite and physically meaningful parts of the energy remain. This vacuum polarization energy is essential for understanding the stability and dynamics of soliton–fermion systems. By generally lowering the total energy of the configuration, it strengthens the soliton’s stability through its interaction with quantum fluctuations of the fermion field.

The VPE of the spinor sector in the ATM model (\ref{atm0}) was previously evaluated for a static, prescribed piecewise-linear pseudoscalar background field \cite{mohammadi}. Using the exact fermionic spectrum of that configuration, the VPE was computed by subtracting the vacuum energies with and without the background. The spinor sector, treated without the scalar field dynamics, has also been investigated for a prescribed sine-Gordon–type soliton background, where numerical simulations and the phase-shift method were used to compute the total Casimir energy \cite{mohammadi1}. In contrast, the present work analyzes the solutions that fully account for the back-reaction of the spinor field on the true soliton of the model, in the variational approach.

In the exactly solvable soliton–fermion system studied here, all normalized negative-energy continuum wave functions in the presence of the soliton, $\varphi(x)$, have been computed explicitly. This allows the VPE to be obtained exactly by directly subtracting the vacuum energy of the system without the soliton from that with the soliton, where the soliton serves as the perturbing background. To examine this more closely, let us follow the approach of \cite{mohammadi, mohammadi1}. So, one has
\br
\nonumber
<\Omega|H|\Omega> - <0|H_{free}|0> &=& \int_{-\infty}^{+\infty} dx  \int_{0}^{+\infty} \frac{dp}{2\pi} (-\sqrt{p^2 + M^2}) \, \zeta^{\star}_{p}  \zeta_p -  \\
&&
\int_{-\infty}^{+\infty} dx  \int_{0}^{+\infty} \frac{dk_1}{2\pi} (-\sqrt{k_1^2 + M^2}) \,  \zeta^{\star\,(free)}_{k_1}  \zeta^{(free)}_{k_1} \label{vpe1} \\
&=&  \int_{0}^{+\infty} dk_1 (-\sqrt{k_1^2 + M^2}) [\hat{\rho}^{(sea)}(k_1) - \hat{\rho}^{(sea)}_{0}(k_1)].\label{vped}
\er
The functions $\zeta_p$ and $\zeta^{(free)}_{k_1}$ ($\zeta \equiv (u,v)^T$) represent normalized wave functions for the negative-energy continuum states in the presence and absence of the soliton, respectively. The factor $ [\hat{\rho}^{(sea)}(k_1) - \hat{\rho}^{(sea)}_{0}(k_1)]$ in Eq. (\ref{vped}) quantifies the spectral deficiency of the continuum states; it is the difference between the densities of negative-energy continuum states with and without the kink

The divergent integrals above have been treated through a formal manipulation: the prescription for subtracting the two divergent integrals in (\ref{vpe1}) is to subtract their integrands at matching values of  $p = k_1$, and then evaluate the remaining $x-$integral. This procedure yields the finite result given in (\ref{vped}).

The VPE is written in (\ref{vpe1})-(\ref{vped}) in the Dirac–sea form 
$E_{vac} = -\sum_{\epsilon_n <0} |\epsilon_n|$, whereas the quantum-field-theory expression is $E_{vac} = -\frac{1}{2}\sum_{n} |\epsilon_n|$. These two representations are generally equivalent only for charge-conjugation-symmetric spectra, which need not hold in general chiral backgrounds. In Appendix \ref{app:chargeconj} we demonstrate that the spectrum of the present model satisfies the corresponding symmetry, i.e. fermion states occur in $\pm E$ pairs, thereby establishing the equivalence of the two forms in our case.

Let us note that indirect techniques-such as the phase-shift method-are sometimes used to compute the VPE in (\ref{vped}). This method relates the momentum derivative of the phase shift to the spectral deficiency of the continuum states. Below, we employ the phase-shift approach to evaluate the VPE in (\ref{vped}). Thus, we have
\br
\label{levinson1}
\frac{1}{\pi} \frac{d}{dk_1}\d(k_1) =
\hat{\rho}(k_1) - \hat{\rho}_{0}(k_1)\er 

Next, using (\ref{levinson1}), the VPE expression in (\ref{vped}) can be rewritten as  
\br
\label{vped0}
VPE &=&<\Omega|H|\Omega> - <0|H_{free}|0> \\
&=& \int_{0}^{+\infty} \frac{dk_1}{\pi} (-\sqrt{k_1^2 + M^2}) \, \frac{d\d(k_1)}{dk_1}  \label{vped1}\\
&=& \int_{0}^{+\infty} \frac{dk_1}{\pi} (-\sqrt{k_1^2 + M^2}) \, \frac{d}{dk_1} (\d(k_1)-\d (+\infty)) \label{vped11}\\
&=&  \int_{0}^{+\infty} \frac{dk_1}{\pi} \frac{k_1}{\sqrt{k_1^2 + M^2}} \, (\d(k_1)-\d(+\infty)) + \frac{M}{\pi} (\d(0)-\d(+\infty))\label{vped2}
\er
Next, we use the common phase shift expression (\ref{phs1}) and (\ref{phs2}) for the spinor components and the Levinson's theorem to rewrite the relevant terms in (\ref{vped2}) as
\br
VPE = - \int_{0}^{+\infty} \frac{dk_1}{\pi} \frac{k_1}{\sqrt{k_1^2 + M^2}} \, \Big[ \arctan{\{\frac{Im[c_1(k_1)]}{Re[c_1(k_1)]}\}} - \arctan{\{\frac{Im[c_1(+\infty)]}{Re[c_1(+\infty)]}\}}\Big] + \frac{M}{2}. \label{vpef}
\er
The last term in (\ref{vpef}) follows from the Levinson's theorem \cite{barton}
\br
\label{lev1}
\d(0)-\d(+\infty) = \pi (n_b - \frac{1}{2}),
\er 
where $n_b$ is the number of bound states in a given scattering channel. In Fig. 4 we show the common phase shift $\d(k) = \d_u(k)=\d_v(k)$ (\ref{phs1}) and (\ref{phs2}) of the scattering states $u$ and $v$ which realizes the Levinson's theorem (\ref{lev1}) for $n_b =1$.   
\begin{figure}
\centering
\label{fig4}
\includegraphics[width=1.5cm,scale=4, angle=0,height=4.5cm]{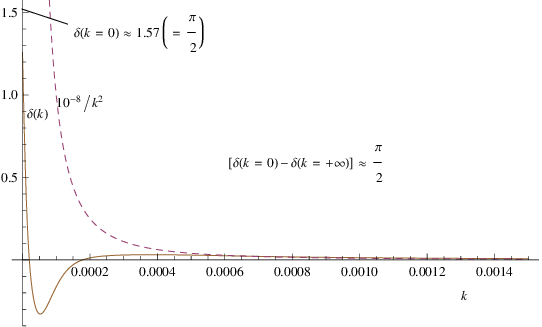}
\parbox{6in}{\caption{(color online) Phase shift $\d(k)$ vs $k$ of the scattering states $u$ and $v$ for  the fermion mass $\,M=2.15 \times 10^{-5}$ and $K= 2 M$. Note that $\d(0)-\d(+\infty) = \frac{\pi}{2}$. The dashed line shows the reference function $\frac{10^{-8}}{k^2}$. Note that the phase shift behaves as  $\frac{1}{k^2}$ in the region $k \rightarrow large$. }}
\end{figure}

A few comments are in order regarding the above calculation of the VPE in comparison with that in Refs. \cite{farhi, farhi2}. First, observe that our model becomes a submodel of the one considered in \cite{farhi, farhi2} when their scalar fields $\phi_{1,2}$ lie on the chiral circle $(\phi_1\,,\,\phi_2) = \frac{1}{2 \hat{\b}} (\cos{2 \hat{\b} \vp}\, ,\, \sin{2 \hat{\b} \vp})$, where $\vp(x \rightarrow -\infty) \rightarrow 0$ and $\vp(x \rightarrow +\infty) \rightarrow \pi/\hat{\b}$, with $\vp$ being the ATM scalar in (\ref{atm0}). Second, in Refs. \cite{farhi, farhi2} the fermion effective energy in the presence of the prescribed classical background was computed numerically within a field-theoretic framework. The standard perturbative renormalization procedure was carried out to one-loop order for the VPE expressions (\ref{vped0})–(\ref{vped1}). Third, their total counterterm contribution to the phase 
shift becomes $\hat{\d}(k) \equiv \frac{8M^2}{k} \int_{0}^{\infty}\, dx  (\vec{\phi}(x)^2-\frac{1}{4 \hat{\b}^2})$. On the chiral circle $\vec{\phi}^2 = \frac{1}{4 \hat{\b}^2}$, when applied to the results obtained here, this counterterm contribution to the VPE vanishes, indicating that the one-loop quantum correction to the energy is finite. Fourth, for scalar configurations lying on the chiral circle, it was observed through numerical analysis in \cite{farhi2} that $\d(k)$ goes like $\frac{1}{k^3}$ for $k$ large. This would imply that the integrand in (\ref{vpef}) decreases more rapidly than $\frac{1}{k^2}$. By contrast, our hybrid analytical and numerical result shows that the integrand decreases as $\frac{1}{k^2}$ for $k$ large. In fact, in Fig. 4 we plot both $\d(k)$ and 
$\frac{1}{k^2}$ to compare their asymptotic behavior for $k \rightarrow large$. Fifth, in (\ref{vpef}), we merge the analytical and numerical components of the spectral method to obtain the one-loop correction consistent with the field-theoretical analyses presented in Refs. \cite{farhi, farhi2}.

In one spatial dimension the scattering problem may be decomposed into even and odd parity channels with phase shifts $\d^{\pm}(k)$. The quantity entering the spectral density and the vacuum polarization energy is the total phase shift $\d_{total} = \d^{+}(k)+\d^{-}(k)$, which equals the phase of the transmission amplitude. Since $t(k)=\frac{1}{c_1}$, the phase shift used here corresponds to this total phase shift, as defined in (\ref{phs1}) and (\ref{phs2}).  

In Appendix \ref{app:phase1} we show that the scattering off the kink is reciprocal in this model, i.e. the left and right reflection amplitudes coincide and the transmission amplitudes differ only by an energy-independent phase. Consequently, the problem admits a single channel-independent phase shift $\d(k)= \arg{(t(k))}= - \arg{(c_1(k))}$, Eqs. (\ref{phs1}) and (\ref{phs2}), which is sufficient for the density-of-states and vacuum-polarization analysis. Moreover, reciprocity implies that the $S-$matrix eigenphases differ only by an energy-independent constant. Therefore, both parity channels share the same energy-dependent phase shift, and the total phase shift entering the density of states can be obtained directly from the transmission amplitude without explicitly performing a parity decomposition. 

For general asymmetric one-dimensional backgrounds the appropriate invariant quantity is the phase of the transmission amplitude \cite{barton}. As discussed in this reference the spectral density and Levinson theorem are governed by the transmission phase $\arg{(t(k))}$, while reflection phases are origin-dependent and a parity decomposition is not applicable. 
 
\subsection{Total energy and stability of the solutions}
\label{sec:totalenergy}

We consider the total energy, which includes contributions from the classical fermion–kink configuration, the valence fermion, and the VPE. The model under study supports both a classical scalar soliton and localized fermionic bound states, arising from the chiral coupling between the scalar and fermionic fields. The topological charge is fixed at 
$Q_{top} =\pm \frac{1}{2}$. Importantly, the scalar sector features a self-interaction potential, in contrast to the model examined in \cite{prd2}, where stable solitons with variable topological charge are generated solely through quantum stabilization mechanisms. Our model constitutes a sub-model of Refs. \cite{farhi, farhi2}, retaining scalar self-coupling, such that the classical solitons obtained in the present paper correspond to solutions of the full theory within an appropriate region of parameter space.

However, upon incorporating quantum effects, the classical treatment must be revisited. In particular, the spatially varying soliton configuration, together with the fermion bound state, should be regarded as minimizing an effective energy that includes both classical contributions and quantum corrections arising from vacuum fluctuations. We seek to compute the total one-loop effective energy of a static configuration $\vp(x)$, assuming that $\vp(x)$  can be described by the width parameter $\sim \frac{1}{2 K}$ of the soliton. For our analytical and numerical calculations, we adopt the ansatz in (\ref{vptau1}) and treat $K$ as a variational parameter, i.e.
\br
\label{vptau11}
\vp^{\pm}(\xi) = -\frac{2}{\hat{\b}} \arctan{\Big[ e^{2 K x}\Big]},\,\,\,\, K \in \IR. 
\er 
Energies are expressed in units of the fermion mass $M$, and distances are written in terms of the dimensionless coordinate $\xi = M x$. In $1+1$ dimensions, both $\vp(x)$ and the coupling $\hat{\b}$ are dimensionless. We therefore rescale $\vp(x)$ by the coupling as $\vp \rightarrow \frac{1}{\hat{\b}} \vp$ so that  $\vp(x \rightarrow -\infty) \rightarrow -\pi$ and $\vp(x \rightarrow +\infty) \rightarrow  0$, and define the dimensionless parameters
\br
\label{zparam}
\zeta = \frac{K}{M},\,\,\,\, \widetilde{m} = \frac{m}{M}.
\er
Using the above rescaling and substituting the soliton (\ref{vptau11}) as an ansatz into the scalar sector of the  energy (\ref{en1}) one can write
\br
E_{cl}[\vp] &=& \frac{M}{\hat{\b}^2} \Big\{2 [1+\frac{\widetilde{m}^2}{4 \zeta^2}] \zeta\Big\},\,\,\,\,\, \widetilde{m} \equiv \frac{m}{M},\\
&\equiv& \frac{M}{\hat{\b}^2} {\cal E}_{cl}(\zeta,\widetilde{m} ).
\er
The fermion one loop contribution to the energy originates from the fermion sector of the energy (\ref{en1}), which arises from the eingenvalues $E_1 = \epsilon$ of the Hamiltonian operator in (\ref{eigene1}), which with the above rescaling becomes multiplied by $M$, i.e. $\epsilon \rightarrow M \epsilon(\zeta)$. So, a  fermion makes a contribution proportional to $M$ and dependent on the variational 
parameter $\zeta$. Therefore, the fermion sector in (\ref{en1}) can be expressed as
\br
E_{F}[\vp] &=& - \frac{1}{2} \sum_b |w_b| + VPE,\\
&\equiv& M \, {\cal E}_{F}(\zeta),
\er
where $w_b$ are the discrete bound state energy levels and $VPE$ was provided in (\ref{vpef}).

In a theory with $N_f$ fermion flavors, quantum corrections from boson loops are suppressed by a factor of $\frac{1}{N_f}$ compared with the fermionic quantum corrections \cite{farhi, farhi2}. So, $N_f$ is introduced as a large-$N_f$ control parameter, making fermion-loop effects dominant and the soliton stabilization mechanism reliably tractable. Then, putting together the classical energy and the one loop energy one gets
\br
\label{totalf}
\frac{E_{tot}[\vp]}{M N_{f}} &= &\frac{1}{ N_f \hat{\b}^2} {\cal E}_{cl}(\zeta,\widetilde{m} ) + {\cal E}_{F}(\zeta) + \frac{1}{N_f} (\mbox{boson loop}) + \mbox{higher loops},\\
&\equiv& {\cal E}_{tot}(K,M,m) + \frac{1}{N_f} (\mbox{boson loop}) + \mbox{higher loops},\label{totalf1}
\er
with
\br
\nonumber
{\cal E}_{tot}(K,M,m) &\equiv & \frac{1}{ N_f \hat{\b}^2} {\cal E}_{cl}(\zeta,\widetilde{m} ) + {\cal E}_{F}(\zeta),\\
&=&\frac{1}{(\hat{\b}\sqrt{N_f})^2} (1+ \frac{\widetilde{m}^2}{4 \zeta^2})(2 \zeta) - 0.986 - \nonumber\\
&&\int_{0}^{+\infty} \frac{dk_1}{\pi} \frac{\frac{k_1}{M}}{\sqrt{k_1^2 + M^2}} \, \Big[ \arctan{\{\frac{Im[c_1(k_1)]}{Re[c_1(k_1)]}\}} - \arctan{\{\frac{Im[c_1(+\infty)]}{Re[c_1(+\infty)]}\}}\Big] + \frac{1}{2},
\label{totalf2}
\er
where the bound-state energy $\frac{w_b}{M} = - 0.986$ in (\ref{bsnumeric}) has been included.
 
In the simultaneous large-$N_f$ and small-$\hat{\b}$, with the quantity $\frac{1}{N_f \hat{\b}^2}$ held fixed, higher order loop contributions are suppressed and can be neglected, leaving only the single fermion loop contribution ${\cal E}_{F}$. Consequently, it suffices to retain the contributions from ${\cal E}_{cl}$ and ${\cal E}_{F}$ in (\ref{totalf}).   

Our next goal is to locate the minimum of the quantity ${\cal E}_{tot}(K,M,m)$ in (\ref{totalf2}) by functionally varying the soliton parameter $K$ for a given set of parameters $\{\frac{1}{\sqrt{N_f} \hat{\b}},\, \widetilde{m}\}$. We choose a value of $\widetilde{m}$ and set the value of $\frac{1}{N_f \hat{\b}^2}$ for small $\hat{\b}$ and large $N_f$.    
\begin{figure}
\centering
\label{fig5}
\includegraphics[width=1.5cm,scale=4, angle=0,height=7cm]{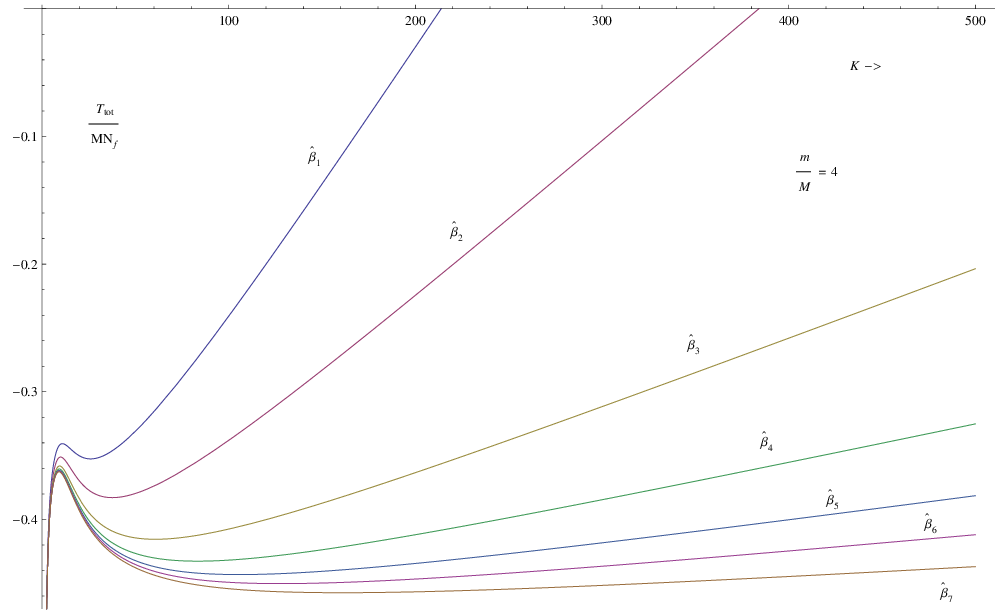}
\parbox{6in}{\caption{(color online) Representatives of the total energy ${\cal E}_{tot} = \frac{E_{tot}}{M N_f}$ (\ref{totalf1}) as a function of variational parameter $K$ for a set $\hat{\b}_1 =3, \hat{\b}_2=4, \hat{\b}_3=6, \hat{\b}_4=8, \hat{\b}_5 = 10, \hat{\b}_6=12,\hat{\b}_7 =15$, $N_f = 100$, and $M = 1, m = 4$. }}
\end{figure}

\begin{figure}
\centering
\label{fig6}
\includegraphics[width=1.5cm,scale=4, angle=0,height=8cm]{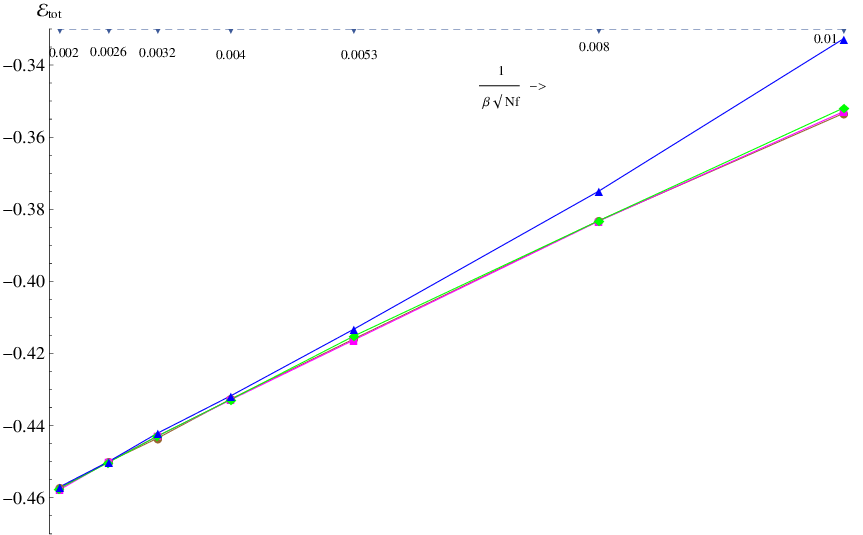}
\parbox{6in}{\caption{(color online) The relative minima of the curves of the type depicted in Fig. 5 defined as ${\cal E}_{tot} = \frac{E_{tot}}{M N_f}|_{min}$, plotted as a function of $\frac{1}{\b \sqrt{N_f}}$,  for $m =2$ (brown), $m=4$ (magenta), $m=8$(green), and $m=32$(blue), respectively. Negative ${\cal E}_{tot}$ indicates binding.}}
\end{figure}

In Fig. 5 we present the results of our numerical calculations for the total energy ${\cal E}_{tot}(K,M,m)$  as a function of the variational soliton parameter $K$, for various values of the coupling constant parameter $\hat{\b}$. We consider the region $K \geq M$, since for $K < M$ the phase shift exhibits a sharp variation, as shown in Fig. 4. For fixed $M=1$, one finds that, for the chosen set of 
$\hat{\b}$  values, $K>1$ is required to ensure a smooth phase shift.

Figure 6 displays the minima $E_{tot}(K,M,m)|_{min} = E_{tot}$ of the curves in Fig. 5 as a function of the composite parameter $\frac{1}{\hat{\b} \sqrt{N_f}}$. The minimum of the total energy identifies a stable soliton. Binding occurs over a broad range of parameters and is intrinsically quantum: it is strongest when $\frac{1}{\hat{\b} \sqrt{N_f}}$, which controls the classical contribution, is small, and weakens as 
$\frac{1}{\hat{\b} \sqrt{N_f}}$ increases. With the appropriate parameter identifications, the resulting behavior is qualitatively similar to that found in Refs. \cite{farhi, farhi2}. 

So, we developed and applied a one-loop variational framework for the renormalizable modified ATM theory in Eq. (\ref{atm0}). Using standard perturbative renormalization, we keep the masses and couplings fixed while minimizing the semiclassical energy over a soliton ansatz. This demonstrates the existence of a quantum-stabilized fermionic soliton in a model that already supports a classical soliton and a zero-mode bound state.
  
\section{Discussions and conclusions}
\label{sec:discuss}

The modified affine Toda model coupled to matter (ATM) (\ref{atm0}) offers a robust framework for examining the interplay between bosonic and fermionic fields, particularly within the setting of deformed integrable theories. In this work, we underscore the pivotal role of the first-order integro-differential system (\ref{sta1})–(\ref{topcurr0}), wherein the scalar sector, Eq. (\ref{topcurr0}), is altered in a manner analogous to the Noether–topological current equivalence characteristic of the undeformed model in its solitonic sector, cf. (\ref{J0vp}). This first-order structure is essential for the analysis of the fermion–soliton configuration energy (\ref{en0})–(\ref{en1}) and for the determination of both bound states and scattering states, which emerge as solutions of the first-order linear systems (\ref{sta1})–(\ref{sta2}) and (\ref{sta11})–(\ref{sta31}), respectively, in the topological soliton background of the model.

Our investigation of the modified ATM model has uncovered a variety of phenomena that substantially advance the characterization of kink–fermion systems. Unlike the Bogomolnyi procedure—which derives first-order field equations by completing the square in the energy functional—our analysis adopts the framework introduced in \cite{prd2}, itself closely aligned with the formulation of \cite{devega}. In that context, the first-order equations for vortices in $2+1$ dimensions arise upon imposing conservation of the energy–momentum tensor, while the approach of \cite{prd2} further rests on the equivalence between the topological and Noether charges. Under these conditions, the energy of the fermion–soliton configuration becomes proportional to the soliton’s topological charge, as expressed in Eq. (\ref{ekf}).

Our results highlight the crucial role of back-reaction effects and demonstrate their impact on the in-gap fermion–kink energy $E_{kf}$ [Eq. (\ref{ekf})], the fermionic bound-state energy $\epsilon$
 [Eq. (\ref{bsnumeric})], and the vacuum-polarization energy given in Eq. (\ref{vpef}).

We have defined the total energy, (\ref{totalf2}), as the sum of the classical fermion-soliton interaction energy $E_{kf}$, the bound-state fermion energy $E_{12}$, and the fermionic vacuum-polarization energy (VPE). Our analysis indicates that the VPE contribution is quantitatively significant when compared to the valence-fermion energy, and therefore must be treated on equal footing with both $E_{kf}$ and $E_{12}$. Furthermore, by examining the full energy functional, we have analyzed the stability of the system under variations of the parameter $K$. As illustrated in Fig. 5, the stability points correspond to the minima of the total-energy profiles $E_{tot}(K,M,m)$ for each set of  dimensionless parameters $\{\widetilde{m}, \frac{1}{\sqrt{N_f} \hat{\b}}\}$. 

We emphasize the central role of the Heun-equation framework in the analysis of both scattering and bound-state spectra. It also provides the location of the virtual states as zeros on the negative imaginary axis, as discussed below Eq. (\ref{c1fr}). While the tau-function formalism provides an efficient means of deriving the zero-mode bound state, it does not accommodate the construction of valence fermion bound states. This limitation stems from the fact that tau functions are analytic over the entire complex $z-$plane, whereas the spinor solutions corresponding to nonzero-energy bound states must be obtained by matching two local solutions of the Heun equation at an intermediate point. Consequently, the matching conditions imposed at 
$x=0$ (\ref{match1})-(\ref{match2}) determine the complex coefficients $c_1(E_1,k)$ and $c_2(E_1,k)$ 
in Eqs. (\ref{c111})–(\ref{c222}), which encapsulate the full scattering data, including the bound-state structure and the phase shifts of the spinor components.

In contrast, the tau-function-based approach employed in \cite{prd2} has been unable to reproduce either the scattering states or the valence fermion bound states in the presence of a sine-Gordon–type background with fixed topological charge $1/2$. We emphasize that the Heun representation is not intended to replace numerical methods but to complement them by providing an analytic framework for the scattering problem. In particular, it makes the analytic structure of the solutions explicit, enables a controlled matching procedure via Wronskians, yields closed expressions for the phase shift, and provides analytic access to bound-state conditions and asymptotic behavior. These features are crucial for implementing the phase-shift representation of the vacuum polarization energy.

Several promising directions for future research remain open. One natural extension involves the study of modified ATM models constructed from higher-rank affine Lie algebras and incorporating scalar potentials as in \cite{jhep22}, which may exhibit richer structural features and enhanced symmetry properties. Incorporating quantum corrections,particularly fermionic vacuum-polarization effects, constitutes another compelling avenue, with the potential to refine the semiclassical picture developed here.

A particularly interesting line of investigation concerns the relationship between the tau-function and Heun-equation frameworks, with the aim of clarifying their respective roles in determining the scattering and bound-state spectra. Additionally, exploring the non-integrable models that emerge due to fermionic back-reaction would provide insights into their behavior within the broader context of quasi-integrability \cite{ferreiraq, np20}. Numerical simulations may also play a crucial role in validating and extending the analytical results, especially in parameter regimes where closed-form solutions are difficult to obtain.

Furthermore, identifying and analyzing candidate experimental platforms capable of emulating the dynamical behavior of the ATM model, such as engineered condensed-matter systems, cold-atom setups, or nonlinear optical lattices, would be highly valuable for benchmarking and validating its theoretical predictions. It is also essential to examine how the stability of the soliton-fermion configurations inherent to the model with self-interacting scalar potential may manifest in broader contexts, including quantum-information processing and condensed-matter physics, where topologically protected states and excitations play a fundamental role.

Finally, the intricate interplay among topology, nonlinear field dynamics, and fermionic degrees of freedom continues to offer a fertile landscape for investigation, with significant potential for uncovering new physical phenomena and enabling future applications.

\[\]
\noindent {\bf Acknowledgements}

The author thanks R. Quica\~no for colaboration in a previous work.    
  
\appendix

\section{Charge conjugation and symmetric spectrum $E \leftrightarrow -E$ }
\label{app:chargeconj}
Here we show the charge conjugation symmetry of the system of equations (\ref{sta11})-(\ref{sta21}). It can be written as 
\br
\label{eigene1}
H \xi &=& E_1 \xi,\\
H &=& \(\begin{array}{cc}
i\pa_x & i M e^{-2 i \b \vp}\\
-i M e^{2 i \b \vp} & -i\pa_x \end{array} \), \,\,\,\,\xi = \(\begin{array}{c}
u  \\
 v\end{array}\).
\er
Notice that the Hamiltonian $H$ for real scalar ﬁeld $\vp$ is a Hermitian operator, i.e. $H^{\dagger} = H$, for $\b$ and $M$ real parameters. In addition, the Hamiltonian satisfies a charge-conjugation (particle–hole) symmetry \cite{jhep22}
\br
\label{ccsym}
\Gamma^{-1} H \Gamma = H^\star, \qquad \Gamma = \pm i\gamma_1 = \pm  \(\begin{array}{cc}
0 & 1\\
1 & 0\end{array} \),
\er
i.e. it is mapped to its complex conjugate by a similarity transformation. Taking the complex conjugate of the eigenvalue equation (\ref{eigene1}) and using the relation (\ref{ccsym}) one gets
\br
H(\Gamma \psi_{E_1}^{\star}) = E_1 (\Gamma \psi_{E_1}^{\star})\,\,\,\rightarrow\,\,\,  \xi = \a_o \Gamma \psi_{E_1}^{\star},\,\,\,\, \a_o = const.
\er
So, $(\Gamma \psi_{E_1}^{\star})$ is also an eigenstate of $H$ (up to a constant factor $\a_o$). Restoring the time dependence of the Dirac field shows that this state is naturally associated with energy $-E_1$, leading to the charge-conjugation relation
\br
\psi_{E_1} = \a_o\,\, \Gamma\,\psi_{-E_1}^{\star}.
\er
Hence the spectrum is symmetric under $(E_1 \leftrightarrow -E_1$), implying identical particle and hole spectra.

A special consequence occurs at $E_1=0$. In this case the conjugate partner has the same energy and the state can satisfy the self-conjugacy (Majorana) condition (set $\a_o =1$)
\br
\psi_0 = \Gamma\,\psi_0^{\star} .
\er
Zero-energy solutions obeying this constraint correspond to Majorana bound states of the system.

So, $E_1 \leftrightarrow -E_1$ symmetry ensures that, even in the kink background, fermionic states occur in $\pm E$ pairs. This allows the vacuum energy to be written as a symmetric sum $-\frac{1}{2} \sum_{n} w_n = \sum_{E_n < 0} E_n\, (w_n = |E_n|)$, so the VPE is obtained by a consistent subtraction of paired spectra. The soliton reshapes the Dirac sea without breaking this symmetry.

\section{Scattering states and definition of a unique phase-shift}
\label{app:phase1}
In this Appendix we demonstrate that scattering off the kink is reciprocal: the left and right reflection amplitudes are identical, while the corresponding transmission amplitudes differ only by an energy-independent phase. As a result, the system is characterized by a single channel-independent phase shift.

We discuss the Eq. (\ref{u1x}) and its relevant scattering amplitudes in order to define the transmission phase shift. Because the governing equation is a second-order differential equation, it admits two linearly independent solutions for each value of $k$. Physically, these independent solutions correspond to scattering states with incidence from opposite directions: one describing a particle incident from the left and the other describing a particle incident from the right. So, let us consider
\br
u_R(x)  & \sim &  \left\{\begin{array}{cc}
   e^{i k x} +  r\, e^{-i k x} & x \rightarrow -\infty, \\
  t \, e^{i k x}   & x \rightarrow +\infty,
\end{array}\right.\label{u1r}\\ 
u_L(x)  &\sim& \left\{\begin{array}{cc}
  t' e^{-i k x} & x \rightarrow -\infty,\\
  e^{-i k x} +  r'\, e^{i k x} & x \rightarrow +\infty. 
\end{array}\right. \label{u1l} 
\er
The function $u_R$ characterizes the asymptotic form of the wavefunction corresponding to a scattering process with incidence from the left, whereas  $u_L$ characterizes the asymptotic region associated with incidence from the right. The coefficients 
$r$ and $t$ denote the reflection and transmission amplitudes, respectively, for left-incident scattering. Analogously, $r'$ and $t'$
 represent the reflection and transmission amplitudes for right-incident scattering. The probability for reflection $R$ and transmission $T$ are given by the usual quantum 
mechanics rule: $R = |r|^2$ and $T = |t|^2$, and similarly for $R'=|r'|^2$ and $T'=|t'|^2$. The particle undergoes either reflection or transmission, and conservation of probability requires that the sum of the corresponding reflection and transmission probabilities is equal to unity, i.e.
\br
\label{consprob}
R + T =1,\,\,\,R' + T' =1.
\er 
General constraints on the transmission and reflection coefficients can be derived by considering the Wronskian $W(u_1,u_2) = u_1 u_2'- u_1' u_2$, where $u_1$ and $u_2$ are distinct solutions of Eq. (\ref{u1x})  corresponding to the same value of $k$. So, one has the next first order differential equation for $W$
\br
\frac{d}{dx} W(x) - 4i K \mbox{sech}(2 K x) W(x) =0,
\er
with solution
\br
\label{w12}
W(x) = W_0 e^{4 i  [\arctan{(2 K x)}+1]}.
\er
So, from (\ref{w12}) on can write $W(-\infty)= W_0$ and $W(+\infty)= W_0 e^{8i}$.  Considering $W(u_R, u_L)$ and equating its values as $x \rightarrow \pm \infty$, i.e. $W(\pm \infty)$, to the relevant values $W(u_R, u_L)(x\rightarrow -\infty)=-2 i k t'$ and $W(u_R, u_L)(x\rightarrow +\infty) = -2 i k t$ one finds 
\br
\label{tt'}
t' = t\, e^{-8 i}.
\er   
Remarkably, the transmission amplitudes $t'$ and $t$ differ only by a constant phase factor. The constant phase shift comes from integrating the Wronskian equation across the soliton.

Next, taking the complex conjugate and the space reflection $x\rightarrow -x$ of (\ref{u1x}) one has 
\br
\label{u1xc}
{u}^{\star\, ''}(-x) - 4i K \mbox{sech}(2 K x) {u}^{\star\, '}(-x) + (E_1^2-M^2 +4 E_1 K \mbox{sech}(2 K x)) u^{\star}(-x) =0.
\er
Notice that (\ref{u1xc}) is the same differential equation as (\ref{u1x}), due to the even function $\mbox{sech}(2 K x)$ under space reflection. So, if  $u_R(x)$ is a solution of (\ref{u1x})  
then so is $u^{\star}_R(-x)$. And, by linearity, so is $[- \frac{1}{r^\star} (t^\star u_R(x) - u_R^{\star}(-x))]$ which is given by
\br
[- \frac{1}{r^\star} (t^\star u_R(x) - u_R^{\star}(-x))] & \sim & 
  \left\{\begin{array}{cc}
-\frac{t^\star r}{r^\star} \, e^{-i k x}   & x \rightarrow -\infty,
\\  
   e^{-i k x} +  \frac{1-|t|^2}{r^\star}\, e^{i k x} & x \rightarrow +\infty \end{array}\right. \label{u1rl}
\er 
This takes the same functional form as (\ref{u1l}) for $u_L$. Then, one has 
\br
\label{rt1}
t' &=& - \frac{t^\star r}{r\star},\\
r' &=& \frac{1-|t|^2}{r^\star}. 
\label{rt2} 
\er
Taking into account $1-|t|^2 = |r|^2$ and (\ref{rt2}) one finds
\br
\label{rr'}
r' = r.
\er 
So, the left/right reflection amplitudes coincide and the transmission amplitudes differ only by a constant phase factor. Comparing the equation (\ref{asympt1}) (multiplied by an overall factor $\frac{1}{c_1}$) with (\ref{u1r}) one gets the relationships 
\br
t &=& \frac{1}{c_1},\,\,\,\,\,\,\, r = \frac{c_2}{c_1} e^{\frac{\pi k}{2M}},\\
\d_{u} &\equiv& \arg{(t)} =  - \arg{(c_1)}. 
\label{phs2app}
\er 
Therefore, it has been established that fermion scattering off the kink is reciprocal and channel-independent: the left and right reflection amplitudes are identical (\ref{rr'}), and the transmission amplitudes differ only by a constant, energy-independent phase (\ref{tt'}). As a result, the scattering problem admits a single well-defined phase shift (\ref{phs2app}) or (\ref{phs1}), ensuring an unambiguous density of states $\hat{\rho}(k_1)$ in (\ref{levinson1}) and a consistent application of Levinson’s theorem (\ref{lev1}). Notice that a constant phase drops out of $\frac{1}{\pi} \frac{d}{dk} \d(k)$ in the l.h.s. of (\ref{levinson1}). These properties are essential for the phase-shift method used in the paper, guaranteeing that the fermionic vacuum polarization energy is uniquely defined and free of directional ambiguities.

\end{document}